\documentclass[10pt]{article}
\usepackage{ametsoc2col}

\usepackage{comment}
\usepackage{graphicx}

\usepackage[labelfont={bf,footnotesize},textfont={footnotesize},margin=1.5em]{caption}

\usepackage{microtype}

\usepackage{amsmath,amsfonts,amssymb,bm}

\usepackage{ulem}

\usepackage{hyperref}
\hypersetup{
	hyperindex,
	breaklinks,
	colorlinks=true,
	linkcolor=blue,
	citecolor=purple,
	bookmarks=true,
	bookmarksopen=true,
	bookmarksopenlevel=2,
	pdfstartpage={1},
	pdfstartview={FitH},
	pdfview={FitH 0},
	pdfauthor={B F Farrell and P J Ioannou},
	pdftitle={Statistical State Dynamics: a new perspective on turbulence in shear flow},
 }
\ifthenelse{\boolean{dc}}
{
\usepackage[all]{hypcap}	
}
{}

\usepackage{natbib}
\usepackage{doi}

\usepackage[usenames,dvipsnames]{xcolor}

\def\C{\bm{\mathsf{C}}}
\def\I{\bm{\mathsf{I}}}

\def\L{\bm{\mathsf{L}}}

\def\S{{\bf S}}

\def\R{{\bf R}}

\def\A{\bm{\mathsf{A}}}
\def\F{\bm{\mathsf{F}}}

\def\Q{\bm{\mathsf{Q}}}

\newcounter{saveeqn}%

\newcommand{\be}{\begin{equation}}
\newcommand{\ee}{\end{equation}}
\newcommand{\bdm}{\begin{equation*}}
\newcommand{\edm}{\end{equation*}}
\newcommand{\bea}{\begin{eqnarray}}
\newcommand{\eea}{\end{eqnarray}}

\newcommand{\partialf}[2]
{
 \ifthenelse{\equal{#1}{}}{\frac{\partial}{\partial #2}}{\frac{\partial #1}{\partial #2}}
}

\renewcommand{\(}{\left(}
\renewcommand{\)}{\right)}

\newcommand{\Del}{\Delta}

\newcommand{\la}{\lambda}

\renewcommand{\b}{\beta}

\renewcommand{\i}{\mathrm{i}}

\newsavebox{\astrutbox}
\sbox{\astrutbox}{\rule[-5pt]{0pt}{20pt}}

\usepackage{upgreek}

\newcommand{\myabstract}{
Traditionally, single realizations of the turbulent state have been the object of study in shear flow turbulence. When a statistical quantity was needed it was obtained from a spatial, temporal or ensemble average of sample realizations of the turbulence. However, there are important advantages to studying the dynamics of the statistical state (the SSD) directly. In highly chaotic systems statistical quantities are often the most useful and the advantage of obtaining these statistics directly from a state variable is obvious. Moreover, quantities such as the probability density function (pdf) are often difficult to obtain accurately by sampling state trajectories even if the pdf is stationary. In the event that the pdf is time dependent, solving directly for the pdf as a state variable is the only alternative. However, perhaps the greatest advantage of the SSD approach is conceptual: adopting this perspective reveals directly the essential cooperative mechanisms among the disparate spatial and temporal scales that underly the turbulent state. While these cooperative mechanisms have distinct manifestation in the dynamics of realizations of turbulence both these cooperative mechanisms and the phenomena associated with them are not amenable to analysis directly through study of realizations as they are through the study of the associated SSD. In this review a selection of example problems in the turbulence of planetary and laboratory flows is examined using recently developed SSD analysis methods in order to illustrate the utility of this approach to the study of turbulence in shear flow.}

\begin{document}

\title{\textbf{\large{Statistical State Dynamics: a new perspective on turbulence in shear flow}}}

\author{\textsc{Brian F. Farrell}\\
\centerline{\textit{\footnotesize{Department of Earth and Planetary Sciences, Harvard University, Cambridge, MA 02138}}}\\
\and
\textsc{Petros J. Ioannou}\thanks{\textit{Corresponding author address:} Petros Ioannou, University of Athens, Department of Physics, Section of Astrophysics, Astronomy and Mechanics, Building IV, Office 32, Panepistimiopolis, 15784 Zografos, Athens, Greece.\newline{E-mail: \href{mailto:pjioannou@phys.uoa.gr}{pjioannou@phys.uoa.gr}}}\\
\textit{\footnotesize{Department of Physics, National and Kapodistrian University of Athens, Athens, Greece}}
}

\ifthenelse{\boolean{dc}}
{
\twocolumn[
\begin{@twocolumnfalse}
\amstitle

\begin{center}
\begin{minipage}{13.0cm}
\begin{abstract}
	\myabstract
	\newline
	\begin{center}
		\rule{38mm}{0.2mm}
	\end{center}
\end{abstract}

\end{minipage}
\end{center}
\end{@twocolumnfalse}
]
}
{
\amstitle
\begin{abstract}
\myabstract
\end{abstract}
\newpage
}



\section{Introduction}

Adopting the perspective of statistical state dynamics (SSD) as an
alternative to the traditional perspective afforded by the dynamics of
sample state realizations has facilitated a number of recent advances in
understanding turbulence in shear flow. SSD reveals the operation in
shear flow turbulence of previously obscured mechanisms, particularly
mechanisms arising from cooperative interaction among disparate scales
in the turbulence. These mechanisms provide physical explanation for
specific phenomena including formation of coherent structures in the
turbulence as well as more general and fundamental insights into the
maintenance and equilibration of the turbulent state.  Moreover, these
cooperative mechanisms and the phenomena associated with them are not
amenable to analysis by the traditional method of studying turbulence
using sample state dynamics. Another advantage of the SSD approach is
that adopting the probability density function (pdf) as a state variable
provides direct access to all the statistics of the turbulence, at least
within the limitations of the approximations applied to the dynamics.
Although the utility of obtaining the pdf directly as a state variable
is obvious, the pdf is often difficult to obtain accurately by sampling
state trajectories even if the pdf is stationary. In the event that the
pdf is time dependent, which is often the case, solving directly for the
pdf as a state variable is the only alternative. While these are all
important advantages afforded by adopting the pdf as a state variable,
the overarching advantage is that adopting the statistical state as the
dynamical variable allows an understanding of turbulence at a deeper
level, which is the level in which the essential cooperative mechanisms
underlying the turbulent state are manifest.

It is well accepted that complex spatially and temporally varying fields
arising in physical systems characterized by chaos and involving
interactions over an extensive range of scales in space and time can be
insightfully analyzed using statistical variables.  In the study of
turbulent systems, examining statistical measures for variables arising
in the turbulence is a common practice; however, it is less common to
adopt statistical state variables for the dynamics of the turbulent
system. The potential of SSD to provide insight into the mechanisms
underlying turbulence has been underexploited in part because obtaining
the dynamics of the statistical state has been assumed to be
prohibitively difficult in practice. An early attempt to use SSD in the
study of turbulence was the formal expansion in cumulants by Hopf
\citep{Hopf-1952,Frisch-1995}. However Hopf's cumulant method was
subsequently restricted in application, in large part due to the
difficulty of obtaining robust closure of the expansion.  Another
familiar example of a theoretical application of SSD to turbulence is
provided by the Fokker-Planck equation which, while very useful
conceptually, is generally intractable for representing complex system
dynamics except under very restrictive circumstances. Because of the
perceived difficulty of implementing SSD to study systems of the type
typified by turbulent flows, the dynamics of these systems has been most
often explored by simulating sample state trajectories which are then
analyzed to obtain an approximation to the assumed statistically steady
probability density function of the turbulent state. This approach fails
to provide insight into phenomena that are associated intrinsically with
the dynamics of the statistical state rather than with the dynamics of
sample realizations. The reason is that while the influence of
multiscale cooperative phenomena on the statistical steady state of
turbulence is apparent from the statistics of sample realizations, the
cooperative phenomena producing these statistical equilibria have
analytical expression only in the SSD of the associated system. It
follows that, in order to gain understanding of turbulent equilibria,
adopting the SSD perspective is essential. But there is a more subtle
insight into the dynamics of turbulence afforded by the perspective of
statistical state dynamics: while the statistical state of a turbulent
system may asymptotically approach a fixed point, in which case the mean
statistics gathered from sample realizations form a valid representation
of the stationary statistical state, the dynamics of the statistical
state may instead be time dependent or even chaotic in which case the
statistics obtained from sample realizations would not in general
correspond to a representation of the statistical state at any time. The
dynamics of such turbulent systems is accessible to analysis only
through its SSD.

Before introducing some illustrative examples of phenomena accessible to
analysis through the use of SSD it is useful to inquire further us to
why this method has not been more widely exploited heretofore. In fact,
as mentioned previously, closure of cumulant expansions to obtain
equilibria of the SSD  has been very extensively studied in association
with isotropic homogeneous turbulence \citep{Kraichnan-1959,
Orszag-1977, Rose-Sulem-1978}. However, the great analytical challenges
posed by this approach and the limited success obtained using it served
to redirect interest toward dimensional analysis and interpretation of
simulations as a more promising approach to understanding the inertial
subrange. It turns out that the inertial subrange, while deceptively
straightforward in dynamical expression, is intrinsically and
essentially nonlinear and as a result presents great obstacles to
analysis. However, the arguably more relevant forms of turbulence, at
least in terms of applications to meteorology, oceanography,
astrophysics, MHD, and engineering fluid dynamics, is turbulence in
shear flow at high Reynolds number. At the core of these manifestations
of turbulence lies a linear dynamics which is revealed by linearization
about the mean flow itself. This linearization in turn uncovers an
underlying simplicity of the dynamics arising from the non-normality of
the linear operator determining the set of dynamically relevant
structures and their interaction with the mean shear. While not
sufficient to eliminate the role of nonlinearity in the dynamics of
turbulence entirely, recognizing the centrality of this non-normality
mediated interaction between perturbations and mean flows motivates a
concept of central importance to understanding turbulence in shear flow
which is the pivotal role of quasi-linear interaction between a
restricted set of non-normal structures and the mean flow. For our
purposes a primary implication of this insight is that the SSD of shear
flow turbulence is amenable to study using closures based on this
interaction.

Consider the large scale jets that are prominent features of planetary
scale turbulence in geophysical flows of which the banded winds of
Jupiter and the Earth's polar jets are familiar examples.  These jets
can be divided into two groups: forced jets and self-sustained jets.
Jupiter's jets are maintained by energy from turbulence excited by
convection arising from heat sources in the planet's interior, and so
the source of the turbulence from which the jets of Jupiter arise may be
regarded as dynamically independent of the jet structure itself, such
jets we refer to as forced. In contrast, the Earth's polar jets are
maintained by turbulence arising from baroclinic growth processes
drawing on the potential energy associated with the meridional
temperature gradient which is directly related to the jet structure by
thermal wind balance. Because baroclinic growth mechanisms depend
strongly on jet structure it follows that the mechanism producing the
jet can not be separated dynamically from the mechanism producing the
turbulence so that these problems must be solved together, such jets we
refer to as self-sustained. In the case of Jupiter's jets the energy
source is known from observations to be convection \citep{Ingersoll-90}
leaving two fundamental problems presented by the existence of these
jets. The first is to explain how the jets arise from the turbulence and
the second how they are equilibrated with the observed structure and
maintained with this structure over time scales long compared to the
explicit time scales of the dynamics.  Although these phenomena manifest
prominently in simulations of sample state realizations, neither is
accessible to analysis using sample state simulation while both have
analytical expression and straightforward solution when expressed using
SSD \citep{Farrell-Ioannou-2003-structural,Farrell-Ioannou-2007-structure}.
Moreover, in the course of solving the SSD problem for jet formation and
equilibration a set of subsidiary results are obtained including
identification of the physical mechanism of jet formation
\citep{Bakas-Ioannou-2013-jas, Bakas-etal-2015}, the structure of the finite amplitude
equilibrated jets, and prediction of the existence of multiple
equilibria jet states \citep{Farrell-Ioannou-2007-structure,Farrell-Ioannou-2009-equatorial,
Parker-Krommes-2013,Constantinou-etal-2014}. But perhaps most
significant is the insight obtained from these SSD equilibria into the
nature of turbulence: the turbulent state is revealed to be
fundamentally determined by cooperative interaction acting directly
between the large energy bearing scales of the mean flow and the small
scales of the turbulence. This fundamental quasi-linearity, which is
revealed by SSD, is a general property of turbulence in shear flow
recognition of which provides both conceptual clarity as well as
analytical tractability to the turbulence closure problem.

An additional issue arises in the case of self--sustained jets that are
maintained by baroclinic growth processes: establishment of the
statistical mean turbulent state for supercritical imposed meridional
temperature gradients requires suppression of the unstable growth. It
has long been remarked that coincident with the equilibration of
baroclinic instability in supercritically forced baroclinic turbulence
is a characteristic organization of the flow into prominent large scale
jets but this association remained an intriguing observation because
although fluctuating approximations to SSD equilibria occur in
realizations, analytic expression of the mechanism of flow instability
equilibration is in general inaccessible within sample state dynamics.
However, this problem can be directly solved using SSD: the equilibrium
turbulent state is obtained in the form of fluctuation-free fixed points
of the autonomous SSD. Moreover, the mechanism of equilibration is also
identified as these fixed points are found to be associated with
stabilization of the supercritical flow  by the jets, in large part by
confinement of the perturbation modes by the meridional jet structure to
sufficiently small meridional scale that the instabilities are no longer
supported, a mechanism previously referred to as the ``barotropic
governor" \citep{Ioannou-Lindzen-1986,James-1987,
Lindzen-1993,Farrell-Ioannou-2008-baroclinic,Farrell-Ioannou-2009-closure}.

In the magnetic plasma confinement problem, which is of great importance
to the quest for a practical fusion power source, the formation of jets
by cooperative interaction with drift wave turbulence is fundamental to
the effectiveness of the plasma confinement \citep{Diamond-2005}. The
jet-mediated high confinement regime is another example of a phenomenon
that arises from cooperative interaction acting directly between large
jet and small perturbation scales in a turbulent flow that can only be
studied directly using SSD.  Drift wave turbulence in plasmas, governed
by the Charney-Hasegawa-Mima or Hasegawa-Wakatani equations, parallels
in dynamics the barotropic or baroclinic turbulence in the Earth's
atmosphere with the Lorenz force playing the role of the Coriolis force
in the plasma case so it is not surprising that the same or very similar
phenomena occur in these two systems. The time dependent statistical
state of drift wave turbulence has natural expression as the trajectory
of the statistical state evolving under its associated SSD
\citep{Farrell-Ioannou-2009-plasmas}. The trajectory of the statistical
state of a turbulent system commonly approaches a fixed point
corresponding to a statistically steady state but in the case of plasma
turbulence the statistical state instead often follows a limit cycle or
even a chaotic trajectory. These time-dependent states are distinct
conceptually from the familiar limit cycles or chaotic trajectories of
sample state realizations. Rather, these statistical state trajectories
represent time-dependence or chaos of the cooperative dynamics of the
turbulence which, while apparent in observation of sample state temporal
variability, has no counterpart in analysis based on sample state
dynamics. An example of a time dependent statistical state trajectory
that is perhaps more familiar to the atmopheric science community is
provided by the limit cycle behavior of the Quasi-biennial Oscillation
in the Earth's equatorial stratosphere
\citep{Farrell-Ioannou-2003-structural}.

A manifestation of turbulence of great practical as well as theoretical
interest is that collectively referred to as wall-bounded shear flow
turbulence, examples of which include pressure driven pipe and channel
flows, flow between differentially moving plane surfaces, over airplane
wings, and in the pressure forced shear flow of the convectively stable
atmospheric boundary layers. These laminar shear flow velocity profiles
have negative curvature and, consistent with the prediction of
Rayleigh's theorem for their inviscid counterparts, these flows do not
support inflectional instabilities. Two fundamental problems are posed
by the turbulence occurring in wall-bounded shear flows: instigation of
the turbulence, referred to as the bypass transition problem, and
maintenance of the turbulent state once it has been established. The
second of these problems, maintenance of turbulence in wall-bounded
shear flow, is commonly associated with what is referred to as the
Self-Sustaining Process (SSP). Transition can occur either by a pathway
intrinsic to the SSD of the background turbulence or alternatively
transition can be induced directly by imposition of a sufficiently large
and properly configured state perturbation \citep{Farrell-Ioannou-2012}.
The former bypass transition mechanism has no analytical counterpart in
sample state dynamics while the latter has been extensively studied
using sample state realizations as for example in \cite{Brandt-2004}. In
contrast, the SSP mechanism maintaining the turbulent state is
fundamentally a quasi-linear multiscale interaction amenable to analysis
using SSD that can be identified with a chaotic trajectory of the SSD
\citep{Farrell-Ioannou-2012}.

In this review one implementation of SSD, referred to as Stochastic
Structural Stability Theory (S3T), will be described. S3T is a second
order cumulant expansion (CE2) closure that employs a stochastic
parameterization to close the expansion. S3T and alternative
implementations of CE2 have been used recently to study: barotropic 
turbulence in planetary atmospheres
\citep{Farrell-Ioannou-2003-structural,Farrell-Ioannou-2007-structure,
Marston-etal-2008,Marston-2010,
Srinivasan-Young-2012, Marston-2012,Bakas-Ioannou-2013-prl, Parker-Krommes-2014-generation, Constantinou-etal-2014, 
Constantinou-etal-2016, Marston-etal-2016, Woillez-Bouchet-2017},
in  equatorial dynamics \citep{Farrell-Ioannou-2009-equatorial,Fitzgerald-Farrell-2017}, 
in baroclinic turbulence \citep{Farrell-Ioannou-2008-baroclinic,  Farrell-Ioannou-2009-closure,Bernstein-Farrell-2010, Farrell-Ioannou-2017-Saturn},
%
the growth of dry convective boundary layer in the atmosphere \citep{Ait-Chaalal-etal-2015}, turbulence in astrophysical flows
\citep{Tobias-etal-2011}, drift-wave turbulence in plasmas
\citep{Farrell-Ioannou-2009-plasmas,Parker-Krommes-2013}, 
and the
turbulence of wall-bounded shear flows \citep{Farrell-Ioannou-2012,Constantinou-etal-Madrid-2014,Thomas-etal-2014,Thomas-etal-2015,Farrell-etal-2016-VLSM,Farrell-Ioannou-2017-bifur,Farrell-etal-2016-PTRSA,Farrell-Ioannou-2017-sync}.

\section{Implementation of SSD: S3T theory and analysis}

S3T implements a closure at second order of the expansion in cumulants
of the system dynamics \citep{Hopf-1952,Frisch-1995}. The Hopf expansion
equations govern the joint evolution of the mean flow (first cumulant)
and the ensemble perturbation statistics (higher order cumulants). A
second order closure is obtained by either a stochastic parameterization
of the terms in the second cumulant equation that involve the third
cumulant \citep{Farrell-Ioannou-1993d,
Farrell-Ioannou-1993e,DelSole-Farrell-1996,DelSole-04} or setting the
third cumulant to zero \citep{Marston-etal-2008}. Restriction of the
dynamics to the first two cumulants is equivalent to either neglecting
or parameterizing by additive noise the perturbation-perturbation
interactions in the fully non-linear dynamics, which removes the
mechanism of the nonlinear perturbation cascade as well as nonlinear
mixing from the dynamics. This closure results in a non-linear,
autonomous dynamical system that governs the evolution of the mean flow
and its associated second order perturbation statistics. The S3T
equations constitute a SSD which, when implemented as the dynamics of a
zonal mean and the covariance of perturbations from this zonal mean
flow, governs the evolution of the statistical state represented by the
zonal mean and a Gaussian approximation to the perturbation covariance
that is consistent with it.

We now review the derivation of the S3T system starting from the discretized Navier-Stokes equations which
can be assumed to take the generic form:
\begin{equation}
  \frac{d x_i}{dt} = \sum_{j,k} a_{ijk} x_j x_k - \sum_j b_{ij} x_j + f_i\ . \label{eq:EQ}
\end{equation}
The flow variable $x_i$ could be the velocity component at the $i$-th
location of the flow and the discrete set of equations~(\ref{eq:EQ})
could arise from discretization of the continuous fluid equations on a
spatial grid. The linear term $\sum_j b_{ij} x_j $ represents
dissipation with $b_{ij}$ a positive definite matrix.  Any externally
imposed body force is specified by $f_i$. In fluid systems $\sum_{i,j,k}
a_{ijk} x_i x_j x_k$ vanishes identically implying in the absence of
dissipation and forcing that $E = \frac1{2} \sum_i x_i^2$ is conserved.

Consider now the averaging operator $\overline{( \,\cdot\, )}$. This averaging operator could be
the zonal mean in a planetary flow, but for now it is left unspecified  but it will be assumed
that it satisfies the Reynolds postulates that require that\\
(a):~$\overline{\lambda \phi(t) + \mu \psi(t) }=\lambda \overline{\phi}(t) +\mu \overline{\psi}(t)$
for all real numbers $\lambda$, $\mu$, and\\
(b):~$\overline{ \overline{\phi}(t) \psi(t)} = \overline{\phi}(t)~\overline{\psi}(t)$.\\
It is also assumed that the averaged quantities are at least as differentiable and integrable as
the original unaveraged fields and that the averaging operator commutes with time translations,
so that
\begin{equation}
\overline{\left ( \frac{\partial \phi}{\partial t} \right )}=  \frac{\partial \overline{\phi}}{\partial t}~,~~~\overline{\left ( \int dt ~ \phi \right )} = \int dt ~ \overline{\phi}~.
\end{equation}
Note that the running time average:
\begin{equation}
\overline{\phi}(t) \equiv \frac{1}{2 T} \int_{t-T}^{t+T} \phi(\tau) d \tau~,
\end{equation}
commutes with time translations, satisfies the linearity assumption~(a),
but does not define an averaging operator as it does not satisfy condition~(b).
These postulates imply  the dependent variable $x_i$ can be decomposed into a mean, and
a perturbation part: $x_i = X_i + x_i'$, where $X_i \equiv \overline{x}_i$, and
 $\overline{x_i'}=0$ and  most importantly that $$\overline{x_i x_j}=
\overline{X_i X_j + X_i x_j' + X_j x_i' + x_i' x_j'} = X_i X_j + \overline{x_i' x_j'}\ ,$$ and that also
\begin{equation}
\overline{ \frac{dX_i}{dt} } = \frac{ d {X_i}} {d t} ~, \overline{ \frac{d x_i'}{dt} } = \frac{ d \overline{{x_i'}}} {d t} = 0~.\end{equation}

In the development that follows we assume that upon decomposing the external forcing  $f_i=F_i +f_i'$
that the $F_i \equiv \overline{f_i}$ are deterministic while the $f_i'$ are stochastic.
Taking the average of \eqref{eq:EQ} we obtain that the mean and perturbation variables evolve according to:
\begin{subequations}
\label{eq:NL}
\begin{align}
&\frac{d X_i}{dt} -\sum_{j,k} a_{ijk} X_j X_k +\sum_j b_{ij} X_j =
 \sum_{j,k} a_{ijk} \overline{x_j' x_k'} + F_i,
 \label{eq:NL_m}\\
&\frac{d x_i'}{dt} = \sum_{j} A_{ij}(X) x_j' + {f_i'}^{\mathcal{NL}}+f_i' \label{eq:NL_p},
\end{align}
\end{subequations}
where
\begin{equation}
A_{ij}(X) = \sum_{k} \left ( a_{ikj} +a_{ijk} \right ) X_k -b_{ij}~,
\label{eq:A}
\end{equation}
and
\begin{equation}
{f_i'}^{\cal NL}=\sum_{j,k} a_{ijk} \left (x_j' x_k'-\overline{x_j'x_k'}\right )~.
\label{eq:F}
\end{equation}
The term $\sum_{j} A_{ij}(X) x_j' $ is bilinear in $X$ and $x'$ and
represents the influence of the mean flow on the perturbation dynamics while the quadratically nonlinear
term ${f'}^{\cal NL}$ represents the perturbation-perturbation interactions which are
responsible for the turbulent cascade in the perturbation variables. The term $F_i^{\cal R} \equiv \sum_{j,k} a_{ijk} \overline{x_j' x_k'}$,
in~\eqref{eq:NL_m} is the perturbation Reynolds stress divergence and
represents the influence of the perturbations on the mean flow.
Equations~\eqref{eq:NL} determine the evolution of the mean flow and the perturbation variables under the full non-linear dynamics~\eqref{eq:EQ} and will be referred to as the NL equations.

The quasi-linear (QL) approximation to NL results when the perturbation-perturbation interactions (given by term ${f'}^{\cal NL}$ in~(\ref{eq:NL_p})) are
either neglected
entirely or replaced by a stochastic parameterization while the influence of perturbations on the mean
is retained fully by incorporating the term $F^{\cal R}$ in the nonlinear mean equation~(\ref{eq:NL_m}). The QL approximation of~(\ref{eq:NL}) 
under the stochastic parameterization ${f_i'}^{\cal NL }+f_i' = \sqrt{\epsilon}\, \sum_{j} f_{ij} dB_{tj}$
is:
\begin{subequations}
\label{eq:QL}
\begin{align}
&\frac{d X_i}{dt} -\sum_{j,k} a_{ijk} X_j X_k +\sum_j b_{ij} X_j =
 \sum_{j,k} a_{ijk} \overline{x_j' x_k'} + F_i ,
 \label{eq:QL_m}\\
& {d x_i'} = \sum_{j=1}^n A_{ij}(X) x_j' dt+ \sqrt{\epsilon}\, f_{ij} \, dB_{tj}. \label{eq:QL_p}
\end{align}
\end{subequations}
The noise terms, $dB_{tj}$, are
independent delta correlated infinitesimal increments of a one-dimensional Brownian motion at time $t$ (cf. \cite{Oksendal})
satisfying:
\begin{equation}
\left < dB_{ti} \right >=0 ~,~ \left < dB_{ti} dB_{sj} \right > = \delta_{ij} \delta(t-s)\, dt~, \label{eq:WN}
\end{equation}
in which $\left <\, \boldsymbol{\cdot} \,\right > $ denotes the ensemble average over realizations of the noise.
Equations~(\ref{eq:QL})
will be referred to as the QL equations.
In the absence of forcing and dissipation the QL equations conserve the
energy $ E^{\cal QL} = \frac1{2}\sum_i \left ( X_i^2 + x_i'^2 \right)$
and, in the presence of bounded deterministic forcing, $F_i$, and dissipation, realizations of the dynamics~(\ref{eq:QL_p}) have
all moments finite at all times.
In general the energy conserved in NL differs from that conserved in
QL and their difference is $E-E^{\cal QL}=\sum_i X_i x_i'$. This cross term vanishes identically if  summation over $i$ provides equivalent action as the chosen averaging operator, otherwise the equality of the energy invariants is true only  on average.
 For example if the averaging operator is the zonal mean and the index refers to the value of the variable on a spatial grid
 the cross term vanishes and then $E=E^{\cal QL}$. We will require that the averaging operators have the property that the QL invariants (energy, enstrophy, etc) are the same as the corresponding NL invariants.

Consider $N$ realizations of the perturbation dynamics~(\ref{eq:QL_p}) evolving under excitation by statistically independent realizations of the forcing but all evolving under the influence of a common mean flow $X$ according to
 \begin{equation}
{d x_i'^r} = \sum_{j=1}^n A_{ij}(X) x_j'^r dt+ \sqrt{\epsilon}\, f_{ij} dB_{tj}^r, ~~~(r=1,\dots,N) .\label{eq:EQL_pa}
\end{equation}
Denote with superscript $r$ the
$r$-th realization so that
$x_j^r(t)$ corresponds to the forcing $dB_{tj}^r$.
Assume further that the mean flow $X$ is evolving
under the influence of the average ${F^{\cal R}}$
over these $N$ realizations, so that:
\begin{align}
\frac{d X_i}{dt} -\sum_{j,k} a_{ijk} X_j X_k +\sum_j b_{ij} X_j =
 \sum_{j,k} a_{ijk} \overline{C_{jk}^N} + F_i~,\nonumber
 \\
 \label{eq:EQL_m}
 \end{align}
 with
 \begin{equation}
 C_{ij}^N = \frac{1}{N} \sum_{r=1}^N x_i'^{r} x_j'^{r}~~,\label{eq:CN}
 \end{equation}
the $N$-ensemble averaged perturbation covariance matrix.
To motivate this ensemble consider it to correspond to the physical situation in which the averaging operator is the zonal mean and
 assume that over a latitude circle the zonal decorrelation scale is such that the latitude circle may be considered
 to be populated by $N$ independent perturbation structures, all of which contribute additively to the Reynolds stresses
 that collectively maintains the zonal mean flow.

 An explicit equation for the evolution of $C_{ij}^N$ is obtained
 using the It\^{o} lemma and~(\ref{eq:EQL_pa}):
 \begin{align}
 d C_{ij}^N & = \frac{1}{N} \sum_{r=1}^N \left ( dx_i'^r x_j'^r + x_i'^r dx_j'^r \right ) +\epsilon \sum_{k=1}^n f_{ik} f_{kj}^T dt \nonumber\\
 &=\sum_{k=1}^n \left ( A_{ik}(X) C_{kj}^N + C_{ik}^N A^T_{kj}(X)+\epsilon f_{ik} f_{kj}^T \right ) dt\, + \nonumber\\
 &\qquad+  \frac{\sqrt{\epsilon}}{N} \sum_{r=1}^N  \sum_{k=1}^n \left (  f_{ik} x_j'^r + f_{jk} x_i'^r \right ) dB_{tk}^r~.
 \label{eq:dC}
 \end{align}
 The stochastic equation~(\ref{eq:dC})  should be understood in the It\^{o} sense, so that
 the variables $x'$ and the noise $dB_{t}$ are uncorrelated in time and the ensemble mean of each of
 $ \left ( f_{ik} x_j'^r + f_{jk} x_i'^r \right ) dB_{tk}^r $ vanishes at all times.
 The corresponding differential equation in the physically relevant Stratonovich interpretation
 is obtained by removing from equation~(\ref{eq:dC}) the term $\epsilon \sum_{k=1}^n f_{ik} f_{kj}^T \, dt$.
 However, both interpretations produce identical covariance evolutions because in the Stratonovich interpretation the mean of $$\sqrt{\epsilon} ( f_{ik} x_j'^r + f_{jk} x_i'^r ) dB_{tk}^r $$
 is nonzero and equal exactly to the term $\epsilon  \sum_{k=1}^n f_{ik} f_{kj}^T dt$ that was removed from the It\^{o} equation.
 This results because the noise in~(\ref{eq:EQL_pa}) enters additively. The noise term
in~(\ref{eq:dC}) can be further reduced using the It\^{o} isometry (cf. \cite{Oksendal}) according to which
any noise of the form $\sum_{k=1}^m g_k(x'_1,\dots,x_n')\, dB_{tk} $
can be replaced by the single noise process $\sqrt{\sum_{k=1}^m g_k^2 (x'_1,\dots,x_n') } ~dB_t$,
in the sense that both processes have the same probability distribution function. Applying the It\^{o} isometry
to the noise terms in~(\ref{eq:dC}) we obtain
 \begin{align}
&\frac{\sqrt{\epsilon}}{N} \sum_{r=1}^N  \sum_{k=1}^n \left (  f_{ik} x_j'^r + f_{jk} x_i'^r \right ) dB_{tk}^r = \nonumber\\
 &~=\sqrt{\frac{\epsilon}{N}}
 \sqrt{ \sum_{k=1}^n  f_{ik} f_{i k} C_{jj}^N +  f_{jk} f_{jk} C_{ii}^N + 2 f_{ik}f_{jk} C_{ij}^N
 }~ dB_t\ , \nonumber
 \end{align}
and~(\ref{eq:dC}) becomes:
 \begin{align}
 d C_{ij}^N &= \sum_{k=1}^n  \left ( A_{ik}(X) C_{kj}^N + C_{ik}^N A^T_{kj}(X)+\epsilon f_{ik} f_{kj}^T \right ) dt\,+ \nonumber \\
&\qquad+ \sum_{k=1}^n \sqrt{\frac{\epsilon}{N}} R_{ik}(C^N) ~d B_{t kj}\ ,  \label{eq:EQL_p}
 \end{align}
 where $d B_{t ij}$ is an $n \times n$ matrix
 of infinitesimal increments of Brownian motion, and the elements of $R_{ij}$ are:
 \begin{equation}
 R_{ij} (C^N) =\sqrt{ {Q}_{ii} C_{jj}^N +Q_{jj} C_{ii}^N + 2 Q_{ij}C_{ij}^N }~.
 \end{equation}
 with $Q_{ij} = \sum_{k=1}^n f_{ik} f_{kj}^T$.
 Equations~(\ref{eq:EQL_m})~and~(\ref{eq:EQL_p}) which govern the evolution of the mean flow interacting with
 $N$ independent perturbation realizations will be referred to as the ensemble quasi-linear
 equations (EQL).

 The stochastic term in the EQL vanishes as the number of realizations increases and in the limit
 $N \rightarrow \infty $ we obtain the autonomous and deterministic system of Stochastic Structural Stability theory (S3T)
 for the mean $X$ and the associated perturbation covariance matrix
 $C_{ij} = \lim_{N\rightarrow \infty} C_{ij}^N$:
 \begin{subequations}
\label{eq:S3T}
 \begin{align}
&\frac{d X_i}{dt} -\sum_{j,k} a_{ijk} X_j X_k +\sum_j b_{ij} X_j =
 \sum_{j,k} a_{ijk} \overline{C_{jk}} + F_i~,\label{eq:S3T_m} \\
 &\frac{d C_{ij}}{d t} = \sum_{k=1}^n  \left ( A_{ik}(X) C_{kj} + C_{ik}A^T_{kj}(X)\right ) +\epsilon \,Q_{ij} ~. \label{eq:S3T_p}
 \end{align}
 \end{subequations}

\section{Remarks on the S3T system }

\begin{enumerate}
\item We have followed a physically based derivation of the S3T equations as in \cite{Farrell-Ioannou-2003-structural}. These equations
can also be obtained using Hopf's functional method \citep{Hopf-1952,Frisch-1995, Marston-etal-2008}.
\item The choice of averaging operator as well as the choice of the stochastic
parameterization for the perturbation-perturbation interaction and the external perturbation forcing used in the S3T model must be
consistent with the dynamics of the turbulent flow being studied.
For example, using the zonal mean as an averaging operator is appropriate when studying jet formation.
\item In the case of homogenous isotropic turbulence the stochastic excitation
must be very carefully fashioned in order to obtain approximately valid statistics using a stochastic closure \citep{Kraichnan-1971r}
while in shear flow the form of the stochastic excitation is not crucial.
The reason is that in shear flow the operator $A_{ij}$ is non-normal
and a restricted set of perturbations participate strongly in the interaction with the mean flow.
As a result the statistical state of the turbulence is primarily determined by the quasilinear interaction
between the mean and these perturbations rather
than by nonlinear interaction among the perturbations.
\item The S3T dynamics exploits the idealization of an infinite ensemble of perturbations interacting with the
mean.  It follows that S3T becomes increasingly accurate as the number of effectively independent perturbations
contributing to influence the mean increases.
\item It is often the case that the zonal mean is an attractive choice for the averaging operator. A stable fixed point of the associated S3T
system then
corresponds to a statistical turbulent state comprising  a mean zonal jet  and fluctuations about it with covariance $C_{ij}$ so that
the pdf of the perturbation field is Gaussian with distribution:
$$p(x_1',\dots,x_n') = \frac{\exp \left[- \tfrac1{2}\sum_{i,j} (\mathbf{C}^{-1})_{ij} x_i' x_j' \right]}{\sqrt{ (2 \pi)^n \det( \mathbf{C})}}\ .$$
\item S3T theory can also be applied to problems in which a temporal rather than a spatial mean is appropriate.
The interpretation of the ensemble mean is then as a Reynolds average over an intermediate time scale, in which interpretation the perturbations are high frequency motions while the mean constitutes the slowly varying flow components
\citep{Bernstein-Farrell-2010, Bakas-Ioannou-2013-prl, Constantinou-etal-2016}.
\item Often the attractor of the S3T dynamics is a fixed point representing a regime with stable statistics.
However, the attractor of the S3T dynamics need not be a fixed point and in many cases a stable periodic orbit emerges as the attracting solution.
In many turbulent systems large scale observables exhibit slow and nearly periodic fluctuation despite short time scales for the
underlying dynamics and the
lack of external forcing to account for the long time scale
(i.e. the Quasi Biennial Oscillation in the Earth's atmosphere,
the solar cycle). S3T provides a mechanism for such phenomena
as reflections of a limit cycle attractor of the ideal S3T dynamics \citep{Farrell-Ioannou-2003-structural}.
\item S3T provides an analytical mechanism for investigating the sensitivity of
the statistical mean state of a turbulent system (the climate as
represented by the model) to perturbations of system parameters.
Small changes in system parameters typically cause correspondingly
small and linearly related variation in the statistical state from which
the sensitivity of the model climate can be inferred.
\item However, at the bifurcation points of the SSD small changes of the system parameters
produce large changes in the statistical state of
the turbulence.
For example, S3T dynamics predicts
bifurcation from the statistical homogeneous regime in which there are no zonal flows
to a statistical regime with zonal flows when a parameter changes or predicts the transition from a regime characterized by two jets to a regime
with a single jet.
\item The EQL equations~(\ref{eq:EQL_m}) and~(\ref{eq:EQL_p}) contain information about the fluctuations remaining in the ensemble dynamics when the number of
ensemble members is finite. These fluctuation statistics can be used to determine the statistics of noise induced transitions between ideal S3T equilibria.
\item The close correspondence of S3T and NL  simulations suggests that turbulence in shear flow
can be essentially understood as determined by quasi-linear interaction occurring directly between a spatial or temporal mean flow and perturbations. This result provides a profound simplification of the dynamics of turbulence, identifies the mechanism determining the statistical mean state in a turbulent flow, and shows that the role of nonlinearity in the dynamics of turbulence is highly restricted.
\item
The S3T system has bounded solutions and if destabilized typically equilibrates to a fixed point which can be identified with statistically stable states of turbulence \citep{Farrell-Ioannou-2003-structural}.
Moreover, these equilibria closely resemble observed statistical states.
For example S3T applied to an unstable baroclinic flow takes the form of baroclinic adjustment that is observed to occur in observations and in simulations \citep{Stone-Nemet-1996, Schneider-Walker-2006, Farrell-Ioannou-2008-baroclinic, Farrell-Ioannou-2009-closure}.
 \end{enumerate}

 \section{Applying S3T to study SSD equilibria and their stability}

S3T dynamics comprises interaction between the mean flow,~$X$, and the
turbulent Reynolds stress obtained from the associated second order covariance of the perturbation field, $C$.
A fixed point of this system, when stable, corresponds to a stationary statistical mean turbulent state.
When rendered unstable by change of a system parameter,
these equilibria predict structural reorganization of the whole turbulent field leading to establishment of a new statistical mean state.
These bifurcations correspond to a new type of instability in turbulent flows associated
with statistical mean state reorganization. Although such reorganizations have been commonly
observed there has not heretofore been a theoretical method
for analyzing or predicting them.

S3T dynamics comprises interaction between the mean flow,~$X$, and the
turbulent Reynolds stress obtained from the associated second order covariance of the perturbation field, $C$.
A fixed point of this system, when stable, corresponds to a stationary statistical mean turbulent state.
When rendered unstable by change of a system parameter,
these equilibria predict structural reorganization of the whole turbulent field leading to establishment of a new statistical mean state.
These bifurcations correspond to a new type of instability in turbulent flows associated
with statistical mean state reorganization. Although such reorganizations have been commonly
observed there has not heretofore been a theoretical method
for analyzing or predicting them.

 We consider first stability of an equilibrium probability distribution function in the context of S3T dynamics.
 The S3T equilibrium is determined jointly by an equilibrium mean flow $X^e$ and a perturbation covariance, $C^e$, that
 together constitute a fixed point of the S3T equations~(\ref{eq:S3T_m}) and~(\ref{eq:S3T_p}):\begin{subequations}
\label{eq:S3Te}
  \begin{align}
&\sum_{j,k} a_{ijk} X^e_j X^e_k -\sum_j b_{ij} X^e_j +\sum_{j,k} a_{ijk} \overline{C^e_{jk}}  + F_i=0\ , \label{eq:S3Te_m} \\
 &\sum_{k}  \left ( A_{ik}(X^e) C_{kj}^e + C_{ik}^e A^T_{kj}(X^e)\right ) +\epsilon Q_{ij} =0\ .
 \label{eq:S3Te_p}
 \end{align}
 \end{subequations}

 The linear stability of a fixed point statistical equilibrium of the S3T system $(X^e,C^e_{ij})$ is determined
from the associated perturbation equations
 \begin{subequations}
\label{eq:dS3T}
 \begin{align}
\frac{d \,\delta X_i}{dt} &= \sum_{k} A_{ik}(X^e) \delta X_k
 +\sum_{j,k} a_{ijk} \overline{\delta C_{jk}} ~, \label{eq:dS3T_m}\\
 \frac{d \,\delta C_{ij}}{d t} &= \sum_{k} \left ( \delta A_{ik} C_{kj}^e + C_{ik}^e \delta A_{kj}^T + \right .\nonumber \\
 &\qquad\qquad\left . + A_{ik}(X^e) \delta C_{kj} +\delta C_{ik} A_{kj}^T(X^e) \right ) ~, \label{eq:dS3T_p}
 \end{align}
 \end{subequations}
 with
 \begin{equation}
\delta A_{ij}= \sum_{k} \left ( a_{ikj} +a_{ijk} \right ) \delta X_k ~.
\label{eq:dA}
\end{equation}

The asymptotic stability of such a fixed point
is determined by assuming solutions of the form $(\widehat{\delta X}_i, \widehat{\delta C}_{ij}) e^{\sigma t}$ with $\delta A_{ij} = \widehat{\delta A}_{ij} e^{\sigma t}$
and by obtaining the eigenvalues, $\sigma$, and the eigenfunctions of the system:\begin{subequations}
\label{eq:sigma}
 \begin{align}
\sigma \widehat{\delta X}_i & = \sum_k A_{ik} (X^e) \widehat{\delta X}_k
+ \sum_{j,k} a_{ijk} \overline{\widehat{ \delta C}_{jk}} \ , \label{eq:sigma__m}\\
 \sigma \widehat{\delta C}_{ij} &= \sum_{k} \left [ \widehat{\delta A} _{ik} C_{kj}^e + C_{ik}^e \widehat{ \delta A}_{kj}^T + \right .\nonumber \\
 &\qquad\qquad\left . + A_{ik}(X^e) \widehat{\delta C}_{kj} +\widehat{\delta C}_{ik} A_{kj}^T(X^e) \right] \ , \label{eq:sigma_p}
 \end{align}
 \end{subequations}

If the attractor of the S3T is a limit cycle then
(\ref{eq:S3T}) 
has time varying periodic solutions $(X_i^p(t), C_{ij}^p(t))$ with period $T$.
The S3T stability of this periodically varying statistical state is determined by obtaining the eigenvalues of the propagator of the time dependent version of
(\ref{eq:dS3T}) over the period $T$.

If the attractor of the S3T is a limit cycle then
(\ref{eq:S3T}) 
has time varying periodic solutions $(X_i^p(t), C_{ij}^p(t))$ with period $T$.
The S3T stability of this periodically varying statistical state is determined by obtaining the eigenvalues of the propagator of the time dependent version of
(\ref{eq:dS3T}) over the period $T$.

\section{Remarks on  S3T instability }
\begin{enumerate}
\item Consider an equilibrium mean flow $X^e$; if $\epsilon$
vanishes identically then $C^e$  also vanishes and S3T stability
theory collapses to the familiar  hydrodynamic instability of this mean flow,
which is governed by the stability of $A(X^e)$. Consequently, S3T instability of $(X^e,0)$ implies the hydrodynamic instability
of $X^e$.
However, if $\epsilon$ does not vanish identically then the instability of the equilibrium state $(X^e,C^e)$
introduces a new type of instability, which is an instability of the collective
interaction between the ensemble mean  statistics of the perturbations and the mean flow.
It is an instability of the SSD and can be formulated only within this framework.
Eigenanalysis of the S3T stability equations~(\ref{eq:dS3T})
provides a full spectrum of eigenfunctions comprising  mean flows and associated covariances that can be ranked according
to their growth rate. These eigenfunctions underly the behavior of QL and NL simulations.
An example of this can be seen in the EQL system, governed by
~(\ref{eq:EQL_m}) and~(\ref{eq:EQL_p}),  which provides noisy reflections of the S3T equilibria and their stability.
 Specifically:  the response of an EQL simulation of a stable S3T equilibrium  manifests structures reflecting
 stochastic excitations of the stable S3T eigenfunctions by the fluctuations in EQL \citep{Constantinou-etal-2014}.
\item Hydrodynamic stability  is determined by
eigenanalysis of the $n \times n$ matrix $A(X^e)$.
However, in order to  determine the  S3T stability of  ($X^e, C^e$) the eigenvalues of the $(n^2 +n)  \times (n^2+n)$   system of equations
(\ref{eq:dS3T}) must be found and special algorithms have been developed for
this calculation \citep{Farrell-Ioannou-2003-structural,Constantinou-etal-2014}.
\item  If $(X^e,C^e)$ is a fixed point of the S3T system,
then   $X^e$ is necessarily hydrodynamically stable, i.e.  $A(X^e)$ is stable.
 This follows because the equilibrium covariance $C^e$  that
 solves~(\ref{eq:S3Te_p}) is determined from the limit
 \begin{equation}
C_{ij}^e = \epsilon \lim_{t \rightarrow \infty} \int_0^t \sum_{k,l}  e^{ A(X^e)(t-s)}_{ik} \,Q_{kl} \,e^{A^T(X^e)(t-s)}_{lj} ds~,
\end{equation}
which does not exist if $A(X^e)$ is  neutrally stable or unstable and consequently in either case $X^e$ is not a
realizable S3T equilibrium.
This argument generalizes to S3T periodic orbits:  if   S3T has a periodic solution, $(X^p(t), C^p(t))$
of period $T$, then
the perturbation operators $A(X^p(t))$ must be  Floquet stable, i.e. the
propagator  over a  period $T$ of $A(X^p(t))$   has eigenvalues, $\lambda$, with $|\lambda|<1$,
so that  the time dependent mean states $X^p(t)$ are hydrodynamically stable.
\item  While S3T stable solutions are necessarily also hydrodynamically stable, the converse is not true:
hydrodynamic stability does not imply S3T stability.  We will give examples  below of hydrodynamically stable flows that
are S3T unstable.
This is important  because it can lead to transitions between turbulent regimes that are the result of
cooperative S3T instability rather than instability of
the associated laminar flow.
\end{enumerate}

\section{Applying S3T to study the SSD of beta-plane turbulence}
\label{section:bar}

\begin{figure}
\centering
\includegraphics[width=19pc,trim = 6mm 0mm 6mm 0mm, clip]{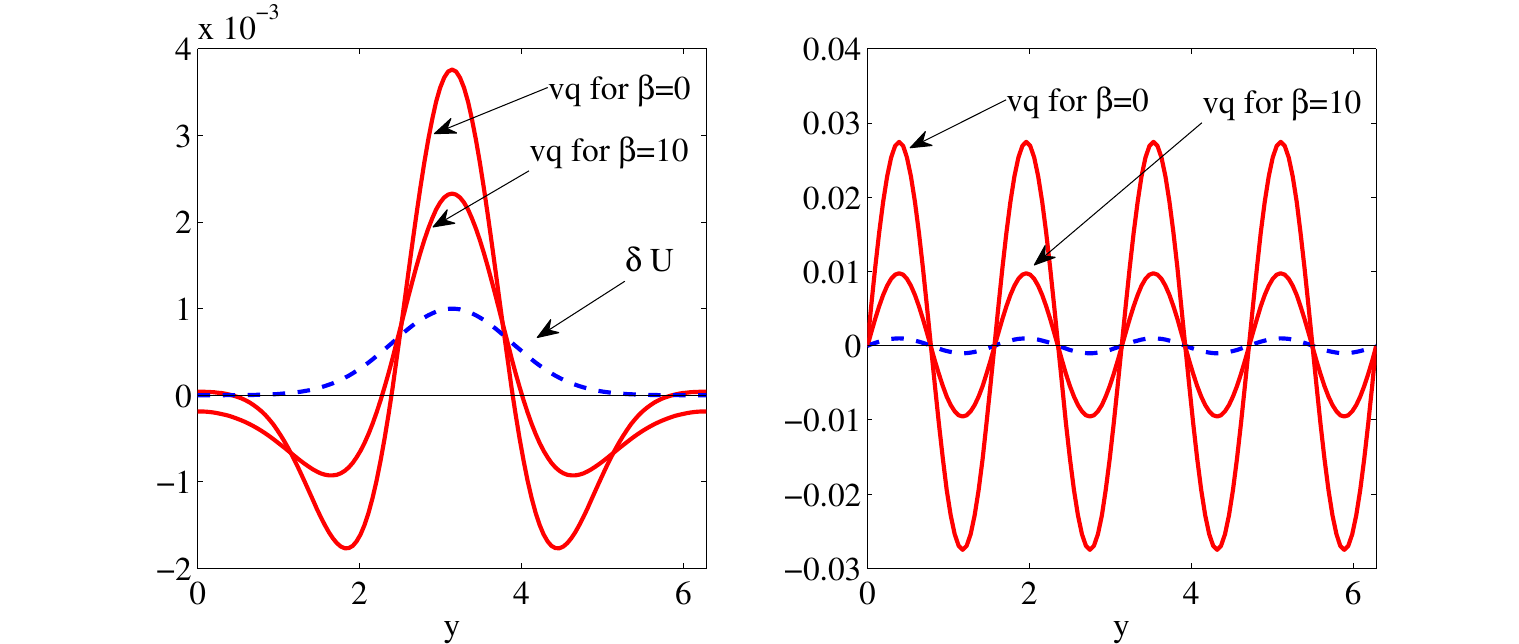}
\caption{
Mean flow acceleration resulting from a small mean flow perturbation imposed on a background
of homogeneous turbulence.
In the absence of a perturbation jet the vorticity flux $\overline{v' q'}=0$. The turbulence is
distorted by the jet perturbation inducing vorticity fluxes that tend to amplify the imposed jet perturbation.
In the left panel is shown a Gaussian jet perturbation together with
the accelerations that it induces for $\beta=0, 10$.
The mean flow acceleration is not the same function as the $\delta U$ that was introduced. Right panel: when
the $\delta U$ is sinusoidal the mean flow acceleration has the same form resulting in exponential growth that leads
to the emergence of large scale jets. Calculations were performed in a doubly periodic
square beta plane box of length $2 \pi$. The coefficient of linear damping is $r=0.1$.
}
\label{fig:test}
\end{figure}

Consider   a barotropic midlatitude beta-plane model for the dynamics of jet formation and maintenance in the Earth's upper troposphere
or in Jupiter's atmosphere at cloud level. For simplicity assume a doubly periodic channel with   $x$ and $y$ Cartesian coordinates along  the zonal
and  the meridional direction respectively. The nondivergent  zonal and meridional velocity fields are expressed  in terms of a streamfunction, $\psi$, as $u=-\partial_y \psi$ and $v=\partial_x \psi$. The planetary vorticity is $2 \Omega+\beta y$, with $\Omega$ the planetary rotation rate and $\beta$ the planetary vorticity gradient evaluated
at the latitude of the midpoint of the channel.
The relative vorticity is
$q = \Delta \psi$ where $\Delta \equiv \partial^2_{xx} + \partial^2_{yy}$ is the Laplacian.
The NL dynamics of this system is governed by the barotropic vorticity equation:
\begin{equation}
 \partial_t q+u\, \partial_x q + v\, \partial_y q + \beta v = {\cal{D}}  + \sqrt{\epsilon}\, {f}~.
 \label{eq:q}
\end{equation}
The term $\cal{D}$ represents linear dissipation with the zonal component of the flow (corresponding to zonal wavenumber $k=0$) dissipated
at the rate $r_m $ while the non-zonal components are dissipated  at rate $r>r_m$ (cf. \cite{Constantinou-etal-2014}).
This dissipation specification allows use of Rayleigh damping while still modeling the physical effect of smaller damping rate at the large jet scale than at the  much smaller perturbation scale.
Periodic boundary conditions are imposed in $x$ and $y$ with periodicity $2 \pi L$.  Distances are nondimensionalized by
$L=5000\;\textrm{km}$ and time by $T=L/U$, where $U=40\;\textrm{m\,s}^{-1}$, so that the time unit is $T=1.5\;\textrm{day}$
and $\beta=10$ corresponds to a midlatitude value.
Turbulence is maintained by stochastic forcing with  spatial and temporal structure, ${f}$, and variance $\epsilon$. 

Choosing as the averaging operator the zonal mean, i.e. $\overline{ \phi} (y,t) = \int_0^{2 \pi} \phi(x,y,t)\, dx / 2 \pi$ and decomposing the fields in zonal mean components and perturbations we obtain the discretized barotropic  QL system:
\begin{subequations}
\label{eq:QLbar}
\begin{align}
\frac{ dU}{dt} &=   \overline{v'q'} - r_m \,U  ~,\label{eq:U_bar}\\
\frac{d q_{k }}{d t}&=   \A_{k  }({U})q_{k }
+ \sqrt{\epsilon}\,   \F_{k} dB_{tk }   ~,\label{eq:q_bar}
\end{align}
\end{subequations}
where  ${ U}$ is the mean flow  state
and the subscript $k=1,\dots,N_k$, in~(\ref{eq:q_bar}) indicates the zonal  wavenumber and
$q_{k}$  the Fourier coefficient of the perturbation vorticity that has been expanded as
$q' = \Re \left ( \sum_{k=1}^{N_k} q_{k} e^{ \i k x} \right ) $.
The  $N_k$ wave numbers include only the zonal wavenumbers that are
excited by the stochastic forcing because  the
$k \ne 0 $ Fourier components do not directly interact in~(\ref{eq:q_bar}) and therefore the
perturbation response is limited to the wavenumbers  directly excited by the stochastic forcing.
The linear operator of the perturbation dynamics~(\ref{eq:q_bar})  is given by:
\begin{align}
\A_{k}({U}) &= - \i k U  - \i k (\beta-D^2U)  \Delta_{k}^{-1}  -
r   ~,\label{eq:Abar}
\end{align}
with $\Delta_k = D^2-k^2$, $\Delta_k^{-1}$ its inverse, and  $D^2=\partial_{yy}$. The continuous operators  are discretized and approximated by matrices.
The  perturbation velocity appearing in
(\ref{eq:U_bar}) is given by
$v'  = \Re \left (    \sum_{k=1}^{N_k} \i k \Delta_{k}^{-1} q_{k }  e^{ \i k x} \right )$,
and  the meridional vorticity flux accelerating the mean flow  is:
\begin{align}
 \overline{v'q'} =    \sum_{k=1}^{N_k}  \frac{k}{2}   {\rm diag } \left [ \Im \left (  \Delta_{k}^{-1} \C_k \right ) \right ]~,
 \label{eq:vq}
\end{align}
with $\Im$ denoting the imaginary part,  $\C_k = q_k q_k^{\dagger}$ the single ensemble member covariance, $\dagger$ the Hermitian transpose,
and $\rm diag $  the diagonal elements of a matrix.
The forcing structure is chosen to be  non-isotropic
with matrix elements:
\begin{align}
{F}_{k i j} & = c_{k}  \left[   e^{-(y_i-y_j)^2/(2s^2)}   +   e^{-(y_i-2 \pi -y_j)^2/(2s^2)} + \right . \nonumber \\
& \qquad\qquad\qquad+\left . e^{-(y_i+2 \pi -y_j)^2/(2 s^2)} \right],
\label{eq:forcing}
\end{align}
with $s=0.2/\sqrt{2}$ and the normalization constants chosen so that energy is injected at each zonal wavenumber $k$ at unit rate.
The delta correlation in time of the excitation ensures that this energy injection rate is the same in the QL and NL simulations
and is independent of the state of the system.  This forcing  is chosen to model  forcing of the barotropic jet in the upper atmosphere
by baroclinic instability. For more
details cf.~\cite{Constantinou-etal-2014}.

Choosing as the averaging operator the zonal mean, i.e. $\overline{ \phi} (y,t) = \int_0^{2 \pi} \phi(x,y,t)\, dx / 2 \pi$ and decomposing the fields in zonal mean components and perturbations we obtain the discretized barotropic  QL system:
\begin{subequations}
\label{eq:QLbar}
\begin{align}
\frac{ dU}{dt} &=   \overline{v'q'} - r_m \,U  ~,\label{eq:U_bar}\\
\frac{d q_{k }}{d t}&=   \A_{k  }({U})q_{k }
+ \sqrt{\epsilon}\,   \F_{k} dB_{tk }   ~,\label{eq:q_bar}
\end{align}
\end{subequations}
where  ${ U}$ is the mean flow  state
and the subscript $k=1,\dots,N_k$, in~(\ref{eq:q_bar}) indicates the zonal  wavenumber and
$q_{k}$  the Fourier coefficient of the perturbation vorticity that has been expanded as
$q' = \Re \left ( \sum_{k=1}^{N_k} q_{k} e^{ \i k x} \right ) $.
The  $N_k$ wave numbers include only the zonal wavenumbers that are
excited by the stochastic forcing because  the
$k \ne 0 $ Fourier components do not directly interact in~(\ref{eq:q_bar}) and therefore the
perturbation response is limited to the wavenumbers  directly excited by the stochastic forcing.
The linear operator of the perturbation dynamics~(\ref{eq:q_bar})  is given by:
\begin{align}
\A_{k}({U}) &= - \i k U  - \i k (\beta-D^2U)  \Delta_{k}^{-1}  -
r   ~,\label{eq:Abar}
\end{align}
with $\Delta_k = D^2-k^2$, $\Delta_k^{-1}$ its inverse, and  $D^2=\partial_{yy}$. The continuous operators  are discretized and approximated by matrices.
The  perturbation velocity appearing in
(\ref{eq:U_bar}) is given by
$v'  = \Re \left (    \sum_{k=1}^{N_k} \i k \Delta_{k}^{-1} q_{k }  e^{ \i k x} \right )$,
and  the meridional vorticity flux accelerating the mean flow  is:
\begin{align}
 \overline{v'q'} =    \sum_{k=1}^{N_k}  \frac{k}{2}   {\rm diag } \left [ \Im \left (  \Delta_{k}^{-1} \C_k \right ) \right ]~,
 \label{eq:vq}
\end{align}
with $\Im$ denoting the imaginary part,  $\C_k = q_k q_k^{\dagger}$ the single ensemble member covariance, $\dagger$ the Hermitian transpose,
and $\rm diag $  the diagonal elements of a matrix.
The forcing structure is chosen to be  non-isotropic
with matrix elements:
\begin{align}
{F}_{k i j} & = c_{k}  \left[   e^{-(y_i-y_j)^2/(2s^2)}   +   e^{-(y_i-2 \pi -y_j)^2/(2s^2)} + \right . \nonumber \\
& \qquad\qquad\qquad+\left . e^{-(y_i+2 \pi -y_j)^2/(2 s^2)} \right],
\label{eq:forcing}
\end{align}
with $s=0.2/\sqrt{2}$ and the normalization constants chosen so that energy is injected at each zonal wavenumber $k$ at unit rate.
The delta correlation in time of the excitation ensures that this energy injection rate is the same in the QL and NL simulations
and is independent of the state of the system.  This forcing  is chosen to model  forcing of the barotropic jet in the upper atmosphere
by baroclinic instability. For more
details cf.~\cite{Constantinou-etal-2014}.

The barotropic S3T system is:
\begin{subequations}
\label{S3Tbar}
 \begin{align}
\frac{d U}{dt} &=   \sum_{k=1}^{N_k} - \frac{k}{2} \Im \left ( {\rm diag} \left ( \Delta_{k  }^{-1} \C_{k  } \right ) \right ) -r_m U~, \label{eq:S3Tbar_m}\\
 \frac{d \C_{k }}{d t}  &=  \A_{k }({ U}) \C_{k }  + \C_{k }\A_{k}^{\dagger}({ U}) +\epsilon\, \Q_{k  } ~,
  \label{eq:S3Tbar_p}
 \end{align}
 \end{subequations}
 with $\C_{k } = \left < q_{k} q_{k }^\dagger \right >$ and $\Q_{k  } =  \F_{k } \F_{k}^{\dagger}$.
 The imaginary part in~(\ref{eq:S3Tbar_m}) requires that we add to the system  an equation
 for the conjugate of the covariance. This is necessary for treating the S3T equations as a dynamical system and for
 analyzing the  stability of S3T equilibria. Alternatively
 we can treat  the real and imaginary part  of the perturbation covariance as separate variables
 to obtain a real S3T dynamical system  as in~(\ref{eq:S3T}).

 Under the assumption that the stochastic forcing  in the periodic channel is homogeneous
~(\ref{eq:S3Tbar_m}) and~(\ref{eq:S3Tbar_p}) admit the homogeneous equilibrium
 \begin{equation}
 {U}^e=0~~~~,~~~\C^ e_{k }  = \frac{\epsilon}{2 r}{\Q_{k }}~,
 \label{eq:Ceq}
 \end{equation}
 as shown in  Appendix A. The assumption of homogeneity of the excitation is crucial for obtaining a
 covariance~(\ref{eq:Ceq})  and associated  turbulent  equilibrium with no flow,
 which requires that the excitation  does not lead to a momentum flux convergence. If
 the excitation were confined to a latitude band, the homogeneous state could not be an S3T equilibrium.
 In that case,   Rossby waves that originate from the region of excitation would dissipate  in the far-field producing
momentum flux convergence into the excitation region   and acceleration of the mean flow  there.
The analysis that follows assumes that the forcing is  homogeneous so that
the homogeneous state~(\ref{eq:Ceq}) is an S3T equilibrium for all parameter values.
The question  is whether  this homogeneous equilibrium is S3T stable. If it becomes S3T
unstable at certain parameter values  this would be an example of a flow that is
hydrodynamically stable but S3T unstable.

\begin{figure}[t]
\centering
\includegraphics[width=19pc,trim = 8mm 1mm 8mm 10mm, clip]{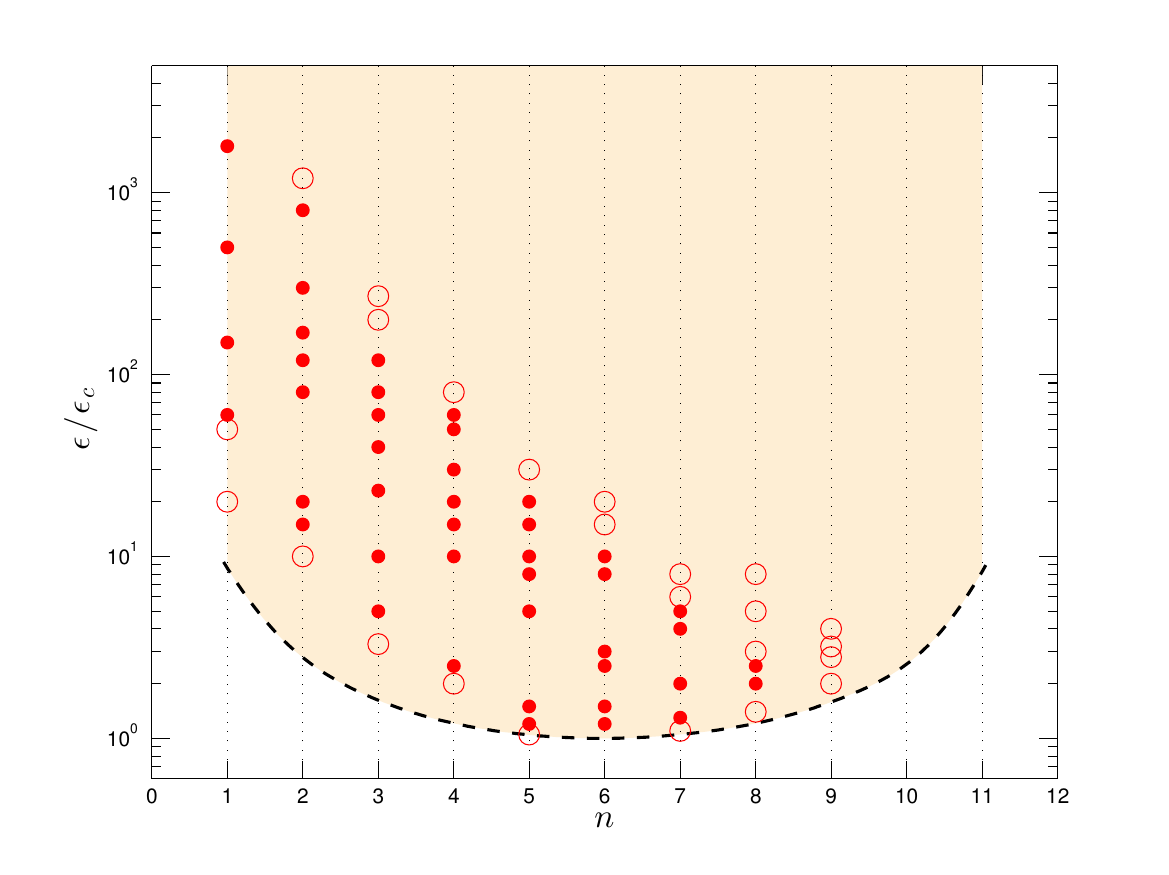}
\caption{S3T stability diagram showing the predicted zonal flow equilibria  as a function of the number of jets, $n$, and
the marginal fractional amplitude of excitation $\epsilon / \epsilon_c$ in the doubly periodic channel (dashed curve).
The amplitude, $\epsilon_c$,  is the minimal excitation amplitude,  obtained from S3T stability analysis,  that
renders the homogeneous state unstable. For parameter values below the dashed curve
the homogeneous state is S3T stable and no jets are
predicted to emerge. 
Above the dashed curve the flow is S3T unstable and
new statistical steady states emerge characterized by the number of finite amplitude  jets across the channel, $n$.
S3T stable  finite amplitude equilibrium jets  are
indicated with a full circle. Note that
for given excitation amplitude there exist multiple S3T stable equilibria characterized by  different numbers of jets.
Unstable S3T equilibria  determined with Newton iterations  are indicated with open circles.
Near the curve of marginal stability the
S3T unstable modes other than the most unstable one at wavenumber $6$ do not continue to
stable finite amplitude structures with the same wavenumber as the instability.
This can be understood as a manifestation of the universal Eckhaus instability as discussed in section 5.2.4 \citep{Parker-Krommes-2014-book} and in \cite{Parker-Krommes-2013}. As the  excitation amplitude increases jet bifurcations occur resulting in the successive establishment of stable S3T equilibria  with smaller $n$. Zonal wavenumbers  $k = 1,\dots,14$ are forced,  $\beta=10$, $r=0.1$, $r_m=0.01$. (Adapted from~\citet{Constantinou-2015-phd}.) }
\label{fig:baloon}
\end{figure}

 \subsection{Formation and structural stability of beta-plane jets}
 \label{section:testbar}

 To motivate our intuition for jet formation by cooperative interaction across spatial scales in beta-plane turbulence
 consider an infinitesimal perturbation of the homogeneous turbulence in the form of a small amplitude jet,  $\delta { U}$, as for example the one shown in
 Fig.~\ref{fig:test}a, and calculate the vorticity fluxes induced  by this perturbation mean flow assuming that the turbulence adjusts  to
 $\delta { U}$ so that it satisfies the equilibrium form of~(\ref{eq:S3Tbar_p})  with zonal mean flow perturbation $\delta U$.
 The modification produced in the turbulence field by this mean flow perturbation produces the steady
 state covariance satisfying the Lyapunov equation: \begin{align}
  \A_{k }(\delta { U}) \C_{k }  + \C_{k }\A_{k }^{\dagger}(\delta {  U})=- \epsilon \Q_{k } ~,
  \label{eq:Cad}
 \end{align}
 for $k=1,\dots,N_k$. By solving~(\ref{eq:Cad})  we
 find that  introducing the infinitesimal jet
 $\delta { U}$ breaks the homogeneity of the turbulence resulting  in an ensemble mean acceleration which can be calculated from~(\ref{eq:vq}).
 This induced acceleration,  which is shown in   Fig.~\ref{fig:test}a, is  upgradient and tends to reinforce the mean flow perturbation that induced it,
 and this occurs even in the absence of   $\beta$ (it vanishes for $\beta=0$ only if the forcing covariance is isotropic).
 Repeated experimentation shows that this positive feedback
occurs for any mean flow perturbation under any  broadband excitation (isotropic or non-isotropic) as long
 as there is power at sufficiently high zonal wavenumbers\footnote{This implies that numerical simulations must be adequately
 resolved for jet formation to occur.}.
 This universal property of reinforcement of preexisting
 mean flow perturbations, revealed through S3T analysis,
 underlies  the ubiquitous phenomenon of emergence of large scale structure in turbulent flows and explains why homogeneous equilibrium states are
 unstable to jet perturbations\footnote{The dynamics leading to this behavior  in barotropic flows  is discussed in \cite{Bakas-Ioannou-2013-jas,Bakas-etal-2015}.
 The counterpart of this process for three dimensional flows is discussed in \cite{Farrell-Ioannou-2012, Farrell-etal-2016-PTRSA,Farrell-Ioannou-2017-bifur}.}.
 It is clear that the acceleration induced by the test perturbation $\delta U$ shown in Fig.~\ref{fig:test}
 is not identical to $\delta U$ and therefore is not an eigenfunction but if
 there exists a mean flow perturbation  that induces accelerations with the same form as  the imposed
 mean flow perturbation then this perturbation would be an eigenfunction of the S3T stability equations
 and it would  grow or decay exponentially without change of form.
S3T  provides the framework for determining systematically the full spectrum of such eigenfunctions enabling full description of
the evolution of a  perturbation of small amplitude near an S3T equilibrium state. In fact, homogeneity in $y$ of the mean state
assures that the mean flow component of the eigenfunctions of the S3T equilibrium~(\ref{eq:Ceq})
are harmonic  so that modulo phase $\delta U_n = \sin ( ny)$, where $n$ indicates the number of jets associated with this eigenfunction.
In Fig.~\ref{fig:test}b is shown a verification  that a single harmonic  induces mean flow acceleration of the same form and is therefore an eigenfunction of the
 S3T stability equations~(\ref{eq:dS3T}). 

\begin{figure}
\centering
\includegraphics[width=19pc]{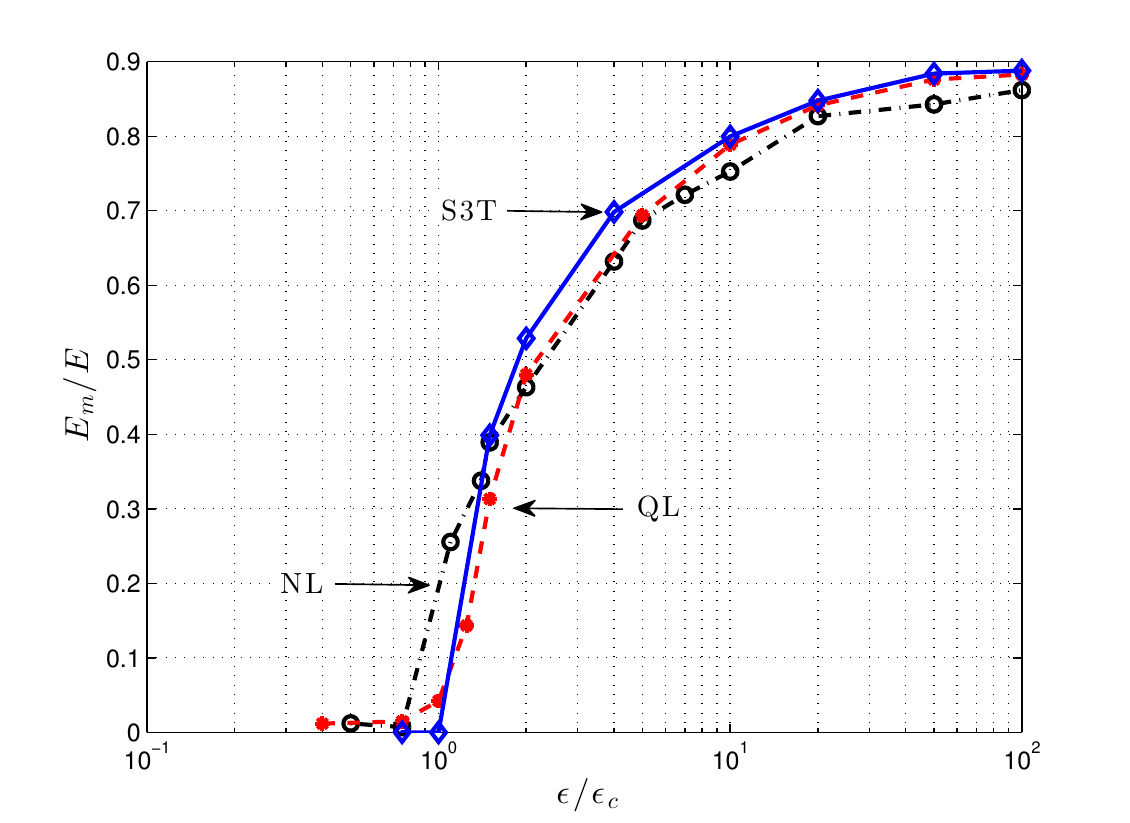}
\caption{The ratio  of the  energy of the zonal mean flow, $E_m$, to the total energy $E$, in S3T (solid and diamonds), QL (dashed and dots), and NL (dash-dot and circles) as a function of forcing amplitude $\epsilon/\epsilon_c$. S3T predicts that the homogeneous flow becomes unstable at $\epsilon/\epsilon_c=1$ with a symmetry breaking bifurcation giving rise to finite amplitude zonal jets. This prediction of S3T is reflected accurately in the sample integrations of both QL and NL. The agreement between  NL and S3T reveals that zonal jet formation is a bifurcation phenomenon and the fact that S3T predicts both the inception of the instability and the finite equilibration of the emergent flows
demonstrates that the essential dynamics of the formation and nonlinear equilibration is captured by QL/S3T. Other parameters as in Fig.~\ref{fig:baloon}. (Adapted from~\citet{Constantinou-etal-2014}.)}\label{fig:bifur}
\end{figure}

\begin{figure}[t]
\centering
\includegraphics[width=19pc,trim = 12mm 0mm 12mm 0mm, clip]{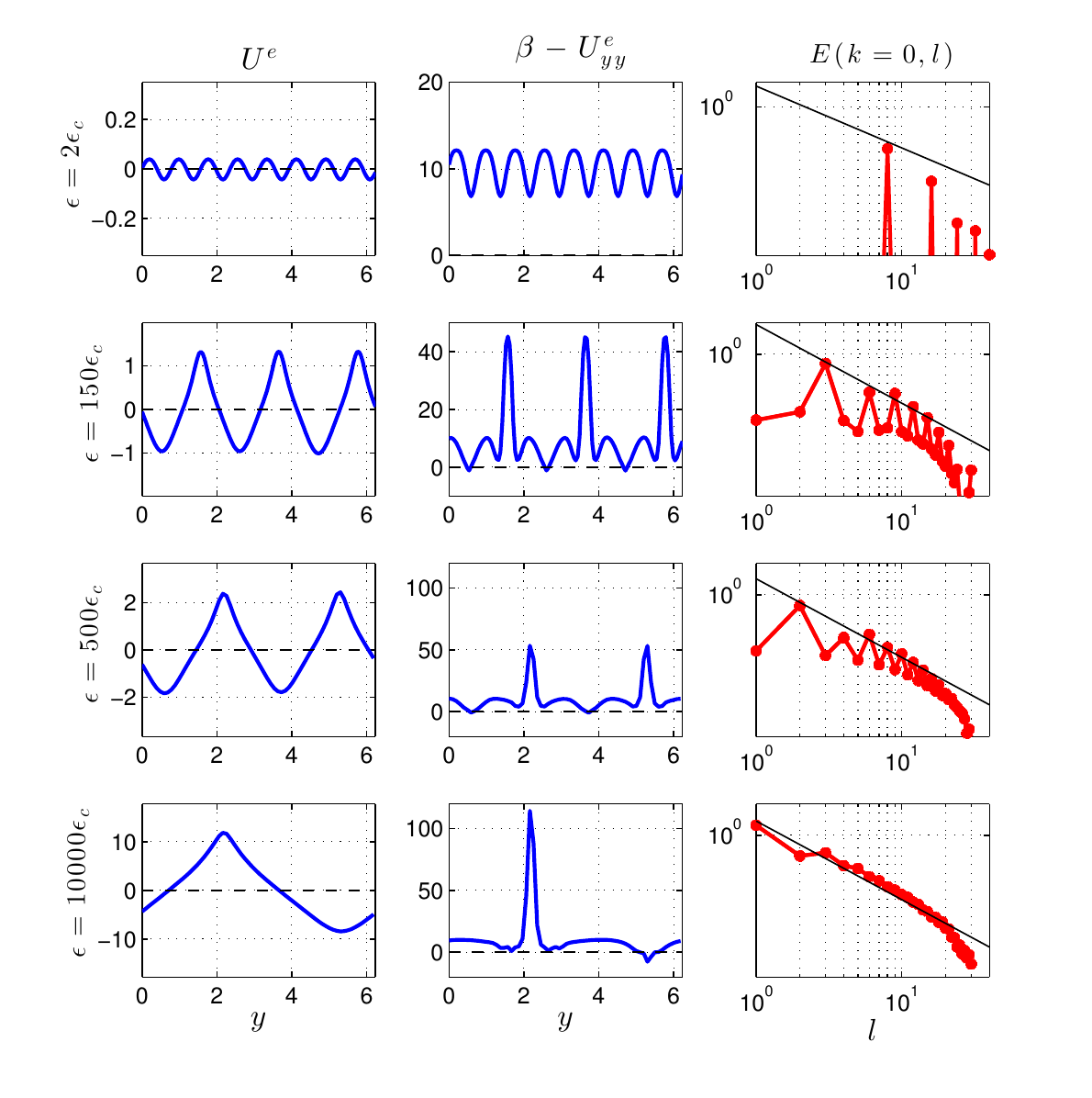}
\caption{Left: The equilibrium mean flow ${U}^e$ for excitation amplitudes $\epsilon/\epsilon_c=2,8,150,500,10^4$.
Center: the corresponding mean vorticity gradient  $\beta-{ U}_{yy}^{e}$. Right: The energy  spectrum of the zonal mean
flow. For the highly supercritical jets the energy spectrum approaches
the approximate $l^{-5}$  dependence  on meridional wavenumber $l$
found in  NL simulations. We argue that in the inviscid limit
this slope should approach  $l^{-4}$ as the prograde jet becomes increasingly sharp.
Other parameters as in Fig.~\ref{fig:baloon}. (Adapted from~\citet{Constantinou-2015-phd}.)}
\label{fig:shape}
\end{figure}

 Consider the stability of the homogeneous equilibrium state~(\ref{eq:Ceq})
 as a function of the excitation amplitude $\epsilon$.
 For $\epsilon=0$ the equilibrium $U^e=0$ and $\C_k^e=0$ is S3T stable.
 For $\epsilon>0$ the universal process of reinforcement of an imposed jet occurs and therefore
 if $r_m=0$ the homogeneous S3T equilibrium would immediately become unstable.
 For $r_m >0$, instability occurs for $\epsilon > \epsilon_c$, where $\epsilon_c$ is the forcing amplitude that renders the S3T stability equation
 neutrally stable.  The normalized critical forcing amplitude for the  S3T eigenfunctions with meridional wavenumber $n=1,\dots ,11 $ is shown in
 Fig.~\ref{fig:baloon}.
 S3T predicts  that for these parameters
 the maximum S3T instability occurs for mean flow perturbations  $\delta U = \sin ( n y)$
  with $n=6$  and thus S3T predicts that the breakdown of the homogeneous state occurs with the emergence
 of  6 jets in this channel.

 \begin{figure}[t]
\centering
\includegraphics[width=19pc]{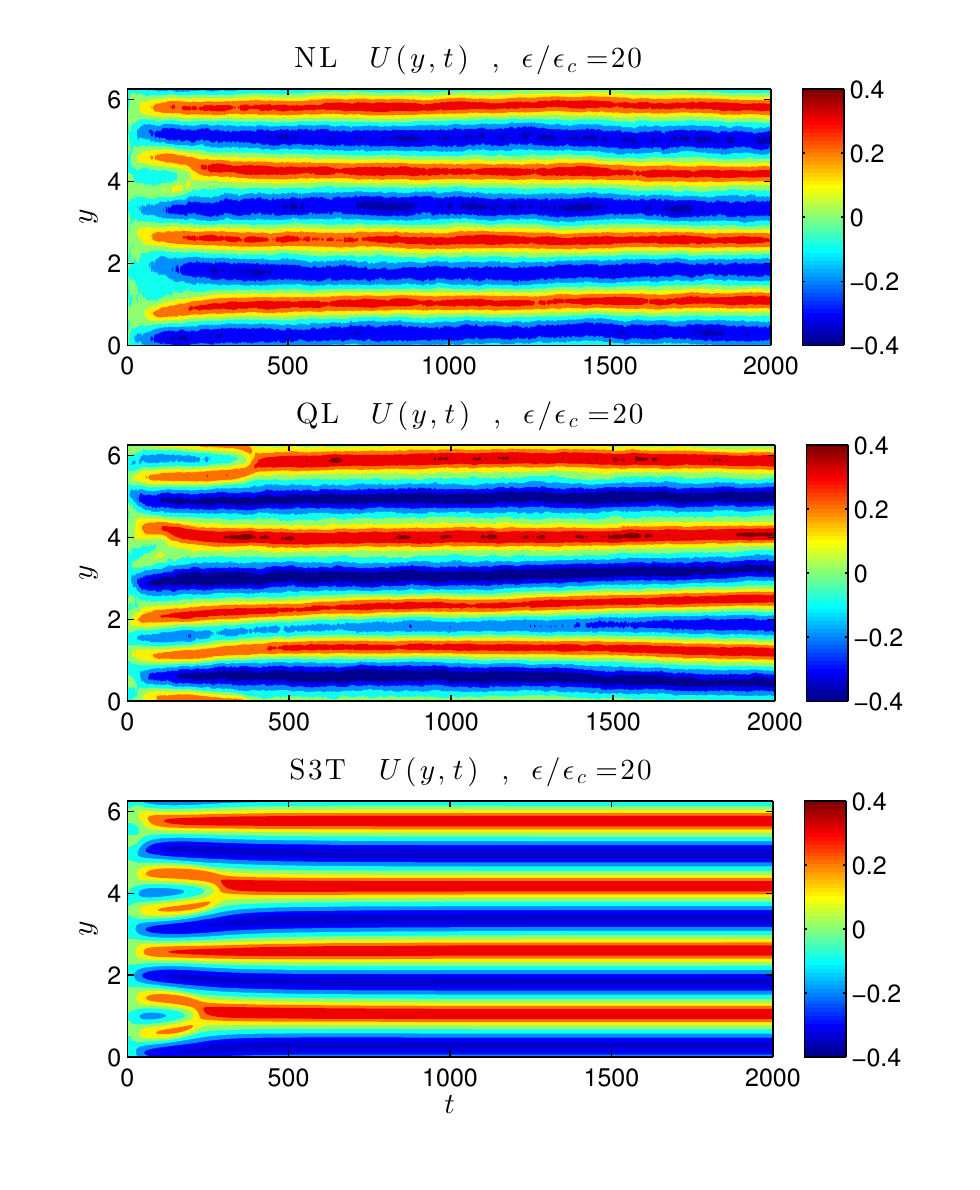}
\vspace{-1em}
\caption{Hovm\"{o}ller diagrams of jet emergence in NL, QL and S3T simulations for  excitation
$\epsilon/\epsilon_c = 20$ as in Fig.~\ref{fig:bifur}.  Shown for the NL, QL and S3T simulations are
$U(y, t)$.  In all simulations the jet structure that first emerges is the $n=6$
maximally growing jet structure predicted by S3T stability analysis.  After a series of mergers S3T is attracted
to a statistical steady state with a  $n=4$ jet. The whole process of jet mergers and S3T equilibration is accurately
reflected in the  QL and NL simulations. This figure shows
that S3T predicts the structure, growth and equilibration of strongly forced jets in
both the QL and NL simulations. Other parameters as in Fig.~\ref{fig:baloon}.}
\label{fig:hov}
\end{figure}

 \subsection{Structure of jets in beta-plane turbulence}

 For parameter values exceeding those required for inception of the S3T instability the S3T attractor comprises statistical
 steady states with finite amplitude jets  that can be characterized
 by   their zonal mean flow index
 $E_m/E$, where $E_m$ is the  kinetic energy density of the mean flow and $E$  to the total  kinetic energy density  of
the  flow, as in \cite{Srinivasan-Young-2012}. A plot of $E_m/E$ as a function of forcing amplitude as predicted by S3T and as observed in QL and NL,
is shown in Fig.~\ref{fig:bifur} and the corresponding
meridional structures of the jet equilibria for various parameter values are shown
 in Fig.~\ref{fig:shape}.  Jets corresponding to S3T equilibria must be   hydrodynamically stable, which  generally requires
 with the mean vorticity gradient $\beta-{ U}^e_{yy}$ of the equilibrium jets not change sign
 as the coefficient of the  linear damping becomes
  vanishingly small. Consistent with being constrained by this criterion, the equilibrated jets at high supercriticality shown in Fig.~\ref{fig:shape}
assume the characteristic shape of sharply pointed prograde jets, with very large negative curvature, and
  smooth retrograde jets which satisfy $\beta \approx {U}^e_{yy}$.  This consideration leads to the prediction   that the mean spacing
  of the jets at high supercriticality is  approximately given by $\sqrt{|U_{\min}|/\beta}$, with $U_{\min}$ the peak retrograde mean zonal velocity.
  At low supercriticality jet amplitude is too low for instability to be a factor in constraining jet structure with typical planetary values of $\beta$
  and the jets equilibrate with nearly the structure of their associated eigenmode,
 as discussed in  \cite{Farrell-Ioannou-2007-structure}. While  the  spacing  of highly supercritical jets   could be interpreted as corresponding
 to Rhines scaling,  the mechanism that produces this scaling
  is associated with the modal stability boundary of the finite amplitude jet and is unrelated to the traditional interpretation of Rhines' scaling in terms of arrest of turbulent cascades.

  S3T also predicts that the  Fourier energy  spectrum of the mean zonal flow
  of the highly supercritical jets has  approximate $l^{-5}$  dependence  on meridional wavenumber,  $l$,
 in accord with highly resolved nonlinear simulations \citep{Sukoriansky-etal-2002,Danilov-04, Galperin-etal-04}
  as well as  observations  \citep{Galperin-etal-2014}.  The finite amplitude jets obtained as fixed  point equilibria in
  S3T  are characterized by near discontinuity
  in the shear at  the maxima of the prograde jets. This discontinuity would be consistent with a zonal energy  spectrum   proportional to    $l^{-4}$. However, this discontinuity can not
  materialize  in the presence of diffusion and it is smoothed resulting in the  approximate  $l^{-5}$  dependence seen in simulations.
  Note that S3T not only predicts
  the energy spectrum of the zonal jets but in addition the  structure and therefore the phase of the spectral components. The occurrence of this power
  law  is theoretically anticipated by S3T because the upgradient fluxes act as a negative diffusion on the mean flow  and as
 a result tend to produce constant shear equilibria at each flank of the jet resulting in the near discontinuity observed  in the
  derivative of the prograde jet  \citep{Farrell-Ioannou-2007-structure}.

 \begin{figure}
\centering
\includegraphics[width=10pc]{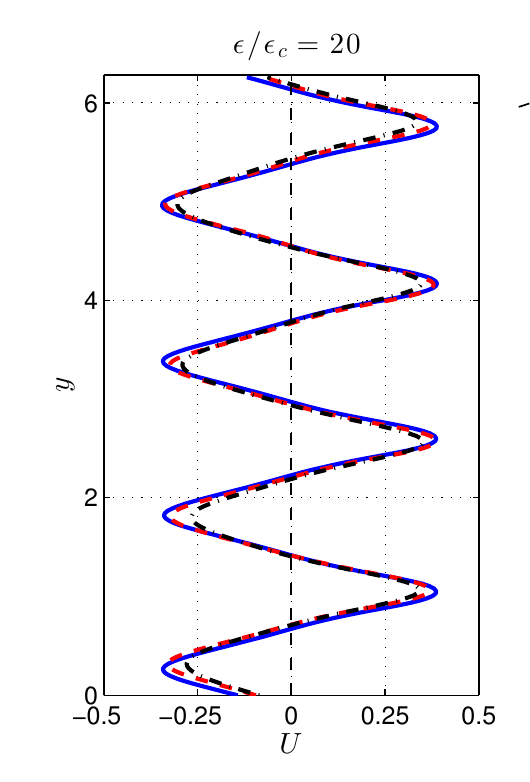}
\caption{The S3T  equilibrium jet (solid-blue) for $\epsilon/\epsilon_c=20$,
and its reflection in a NL simulation (dash-dot-black) and a QL
simulation (dashed - red). The jets in NL and QL undergo small fluctuations as is evident
in Fig.~\ref{fig:hov} so shown is the average  jet structure over 500 units of time.
This figure shows that
S3T predicts the structure  of strongly  forced jets in
both the QL and NL simulations. Other parameters as in Fig.~\ref{fig:baloon}.}
\label{fig:meanU}
\end{figure}

\subsection{Comments on statistical equilibria in beta-plane turbulence}

turbulence
are predictions of S3T that could not result from analysis within
the NL and QL systems
as  in both NL and QL  turbulent states with stable statistics do not exist as fixed point equilibria so that
a meaningful structural stability  analysis of statistically stable equilibria states can not  be performed.
However, as revealed in Fig.~\ref{fig:bifur}, Fig.~\ref{fig:hov} and  Fig.~\ref{fig:meanU},
reflections of the bifurcation structure and of the finite amplitude equilibria predicted by the S3T closure dynamics
can  be  clearly seen both in QL and NL sample simulations.
The reflection in NL of the bifurcation structure predicted by S3T for $\epsilon \approx \epsilon_c$
is particularly significant because it  shows that the S3T perturbation instability faithfully reflects  the physics underlying
jet formation in turbulence even as it manifests for  infinitesimal mean flows. For a more detailed discussion
cf.~\cite{Constantinou-etal-2014}.

\section{Applying S3T to study SSD equilibria in baroclinic turbulence}

A 2D barotropic fluid lacks a source term to maintain vorticity against dissipation and therefore
it cannot self-sustain turbulence. Perhaps the simplest model that self-sustains turbulence is
the baroclinic two-layer model, which is essentially
two barotropic fluids sharing a horizontal boundary. We use this model to study the statistical state dynamics of baroclinic turbulence.
The perturbation-perturbation nonlinearity has been shown to be accurately parameterized in S3T studies of
baroclinic turbulence by using a state independent stochastic excitation as a closure \citep{DelSole-Farrell-1996,DelSole-Hou-1999,DelSole-04}.
The dynamics of baroclinic turbulence has also been studied using S3T with
a state dependent closure with similar results  \citep{Farrell-Ioannou-2009-closure}.

 Consider a two layer fluid with quasi-geostrophic dynamics. The layers are of equal depth $h/2$ and
 bounded by horizontal rigid walls at the bottom and the top. Variables in the top layer are denoted with
subscript 1, and in the bottom layer with subscript 2.
The top layer density is $\rho_1$ and the bottom $\rho_2$ with $\rho_2 > \rho_1$
and the potential vorticity of the layer is $q_i = \Del \psi_i +\beta y +(-1)^i 2\lambda^2(\psi_1-\psi_2)/2$ ($i=1,2$),
with $\lambda^2= 2 f_0^2/ ( g' h ) $, where  $g'=g(\varrho_2-\varrho_1)/\varrho_1$  is the reduced gravity,
and $f_0$ the Coriolis parameter so that, in terms of the Rossby radius of deformation $L_d = \sqrt{g' h} / f_0 $,
$\lambda = \sqrt{2}/L_d$.
The flow is relaxed to a constant temperature gradient in the meridional direction ($y$) which through the thermal wind relation
induces, in the absence of turbulence and at equilibrium, a meridionally independent mean shear $H_T$ in the zonal ($x$) mean velocity.
This shear is taken without loss of generality to result from relaxation to zero velocity in the bottom layer ($2$) and
a constant mean flow with stream function $-H_T y$ in the top layer (1). Relaxation to this velocity structure is induced by
Newtonian cooling and the bottom layer is dissipated by Ekman damping. The quasi-geostrophic dynamics governing this
system is given by:\begin{subequations}
\label{eq:BRCL}
\begin{align}
\partial_t q_1 + J(\psi_1,q_1) &=\;\;\, 2\la^2 r_T \frac{\psi_1-\psi_2 +H_T y}{2} \ ,\\
\partial_t q_2 + J(\psi_2,q_2) &= -2\la^2 r_T \frac{\psi_1-\psi_2 +H_T y}{2} - r \Del\psi_2\ ,
\end{align}\end{subequations}
in which the advection of potential vorticity is expressed using the
Jacobian $J(\psi,q)=\partial_x \psi \partial_y q - \partial_y \psi \partial_x q$,
$r_T$ is the coefficient of the Newtonian cooling and $r$ is the coefficient of Ekman damping.
Equations~(\ref{eq:BRCL}) have been made non-dimensional with length scale $L_d=1000~\rm km$ and time scale $1~\rm day$.
With this scaling $\lambda=\sqrt{2}$, the velocity unit is $11.5~\rm m s^{-1}$ and a typical midlatitude value of  $\beta = 1.4$.
These equations can be expressed alternatively in terms of the barotropic  $\psi = ( \psi_1+\psi_2)/2$ and
baroclinic $\theta = (\psi_1-\psi_2)/2$ streamfunctions as:\begin{subequations}
\begin{align}
\partial_t \Del \psi &+ J(\psi,\Del\psi) + J(\theta,\Del\theta) + \beta\psi_x = -\frac{r}{2}\Del(\psi-\theta) \\
\partial_t \Del_\la \theta &+ J(\psi,\Del_\lambda\theta) + J(\theta,\Del\psi) + \beta \theta_x = \nonumber \\
& \qquad=\frac{r}{2}\Del(\psi-\theta) +2\lambda^2 r_T\(\theta+\frac{H_T}{2}y \)\end{align}\end{subequations}
where $\Del_\lambda\equiv\Del -2\lambda^2$.

The barotropic and baroclinic streamfunctions are decomposed into a zonal mean (denoted with capitals) and deviations from
the zonal mean (referred to as perturbations) as:
\begin{equation}
\psi=\Psi + \psi'~~~\ , ~~~~~\theta= \Theta+\theta'\ ,
\end{equation}
and denote with $U=-\partial_y \Psi$ the barotropic zonal mean flow
and with $H=-\partial_y \Theta$ the baroclinic zonal mean flow.
With this decomposition we obtain the equations for the evolution of the barotropic and baroclinic zonal mean flows:
\begin{subequations}
\label{eq:NL_mbrcl}
\begin{align}
\partial_t U &=  \overline{ \psi_x' \psi_{yy}' + \theta_x' \theta_{yy}' }-\frac{r}{2} ( U- H) -r_m U + \nu D^2 U\\
\partial_t D^2_\la H  &=D^2 \left ( \overline{\psi_x' D^2_\la \theta' +  \theta_x' \psi_{yy}' } \right )+ \frac{r}{2} D^2 ( U- H)+\nonumber\\
&~~~~~~~~~ + 2\la^2 r_T \left (
H -\frac{H_T}{2} \right ) -r_m D^2_\la H +\nu D^2 D^2 H\ ,
\end{align}\end{subequations}
with $D^2 \equiv \partial_{y}$, $D^2_\la \equiv \partial_{y}-2\la^2$, subscripts $x$ and $y$ denoting differentiation
and the overline denoting zonal averaging.
The corresponding barotropic and baroclinic components of the zonal mean flow deviations are
$U = -\Psi_y $ and $H= - \Theta_y$. The equations for the evolution of the perturbations are:
\begin{subequations}
\label{eq:NL_pbrcl}
\begin{align}
&\partial_t \Del \psi' + U  \partial_x\Del\psi' + H \partial_x\Del\theta' + (\b - D^2 U ) \partial_x \psi' - \nonumber \\
&~~-D^2 H \partial_x \theta'= -\frac{r}{2}\Del(\psi'-\theta') -r_p \Del \psi' +\nu \Del \Del \psi' \nonumber\\
&\qquad\qquad\qquad\qquad- J(\psi',\Del \psi' )' -J (\theta', \Del \theta')' ,\label{eq:psi'}\\
&\partial_t \Del_\la \theta' + H  \partial_x\Del\psi' + U \partial_x\Del_\la \theta'
+ (\b - D^2 U) \partial_x \theta' -\nonumber \\
&- D^2_\la H \partial_x \psi' =
\frac{r}{2}\Del(\psi'-\theta') +2\la^2r_T\theta' -r_p \Del_\la \theta' +\nonumber\\
&\qquad\qquad\qquad\qquad+\nu \Del \Del \theta'- J(\psi',\Del_\la \theta' )' -J (\theta', \Del \psi')'\ ,\label{eq:theta'}
\end{align}
\end{subequations}
with the prime Jacobians denoting the perturbation-perturbation interactions,
\begin{equation}
J(A,B)'=J(A,B)-\overline{J(A,B)}\ .
\end{equation}
We have allowed in~(\ref{eq:NL_mbrcl}) and~(\ref{eq:NL_pbrcl}) for linear dissipation of the mean at rate $r_m$
and of the perturbations at rate $r_p$.
This choice in decay rates reflects the different rates of dissipation
of the mean flow and the perturbation field, which in natural flows is concentrated at smaller scales.
We have also included diffusive dissipation of the velocity field with coefficient $\nu$.
Equations~(\ref{eq:NL_mbrcl}) and~(\ref{eq:NL_pbrcl}) comprise the NL system that governs the two layer baroclinic flow.
We impose periodic boundary conditions at the channel walls on $\Psi$, $\Theta$, $\psi'$, $\theta'$  as in \cite{Haidvogel-Held-1980,Panetta-93}.
These boundary conditions can be verified to require that the zonally and meridionally averaged velocity at all times remains equal to that of the radiative
equilibrium flow, which in turn implies that the temperature difference between the channel walls,
and therefore the channel mean criticality, remains fixed.

The corresponding QL system is obtained
by substituting for the perturbation-perturbation interaction a state independent
and temporally delta correlated stochastic excitation together with sufficient added diffusive dissipation to obtain
an approximately energy conserving closure \citep{DelSole-Farrell-1996,DelSole-Hou-1999,DelSole-04}. Under
these assumptions the QL perturbation equations
for the Fourier components of the barotropic and baroclinic streamfunction are:
\begin{subequations}
\label{eq:QLbrcl}
\begin{align}
& \frac{d\psi_k}{dt} = \A_k^{\psi \psi} \psi_k + \A_k^{\psi \theta} \theta_k + \sqrt{\epsilon} \,\Delta_k^{-1} \F_k\,\xi^{\psi} (t) \ ,\\
& \frac{d \theta_k}{dt} = \A_k^{\theta \psi} \psi_k + \A_k^{\theta \theta} \theta_k + \sqrt{\epsilon} \,\Delta_{k \la}^{-1} \F_k \,\xi^{\theta} (t) \ ,
\end{align}
\end{subequations}
 in which  we have assumed that the barotropic and baroclinic streamfunctions are excited respectively
 by
 $\sqrt{\epsilon } \Delta_k^{-1} \F_k \xi^{\psi}$ and $\sqrt{\epsilon} \Delta_{k \la}^{-1} \F_k \xi^{\theta}$.
 We also assume that $\xi^{\psi}$ and $\xi^{\theta}$ are
 independent temporally delta correlated stochastic processes of unit variance and that the
 perturbation fields have been expanded as $\psi'= \Re \( \sum_k \psi_{k} e^{i k x} \)$ and
 $\theta'= \Re \( \sum_k \theta_{k} e^{i k x} \)$.
 The operators $\A_k$ linearized about the mean zonal flow ${\cal U }=[U,H]^T$ are
\begin{equation}
\A_k({\cal U})~=~
\left(%
\begin{array}{cc}
\A_k^{\psi \theta} & \A_k^{\psi \theta} \\
 \A_k^{\theta \psi} & \A_k^{\theta \theta} \\
\end{array}%
\right)
\end{equation}
 with  :
 \begin{subequations}
\label{eq:Abrcl}
\begin{align}
 \A_k^{\psi \psi}&=\Delta_k^{-1} \left [ -\i k  U
 \Delta_k -\i k \left (
\beta - D^2 U \right ) \right ] -\frac{r}{2} -r_p~+\nu \Delta_k, \\
 \A_k^{\psi \theta}&= \Delta_k^{-1} \left [ -\i k  H \Delta_{k } +\i k D^2 H
\right ]+\frac{r}{2} \ , \\
 \A_k^{\theta \psi}& = \Delta_{k \la}^{-1} \left [ -\i k  H \Delta_k
 +\i k D^2_\la H  +
  \frac{r}{2}\Delta_k \right ] \ , \\
 \A_k^{\theta \theta}& = \Delta_{k \la}^{-1} \left [ -\i k U \Delta_{k \la}
 -\i k \left ( \beta - D^2 U \right ) - \right.\nonumber\\
 &\hspace{2.7cm}\left. \vphantom{\left( \beta - D^2 U \right )} -\frac{r}{2} \Delta_k + 2 r_T \lambda^2 +\nu \Delta_k \Delta_k \right ]-r_p\ ,
\end{align}\end{subequations}
and $\Delta_k \equiv D^2- k^2$, $\Delta_{k \la} \equiv \Delta_k - 2 \lambda^2$. Continuous operators are discretized and the dynamical
operators approximated by finite dimensional matrices. The
states $\psi_k$ and $\theta_k$ are represented by a column vector with entries
the complex value of the barotropic and baroclinic streamfunction at the collocation points in $(y)$.

The corresponding S3T system is obtained by forming from the QL equations~(\ref{eq:QLbrcl})
the Lyapunov equation for the evolution for the
zonally averaged covariance of the perturbation field which takes the form:
\begin{equation}
 \frac{d \C_k }{d t}~ =~\A_k({\cal U}) \, \C_k~+\C_k \,\A_k^{\dagger}({\cal U})~+~\epsilon \Q_k\ ,
  \label{eq:ensemblea}
 \end{equation}
 with the covariance for the wavenumber $k$ zonal Fourier component defined as:
\begin{equation}
\C_k ~= ~
\left(%
\begin{array}{cc}
\C_k^{\psi \psi} & \C_k^{\psi \theta} \\
\C_k^{\psi \theta \dagger} & \C_k^{\theta \theta} \\
\end{array}%
\right)\ ,
\end{equation}
where $\C_k^{\psi \psi}= \langle \psi_k \psi_k^{ \dagger}\rangle $,
 $\C_k^{\psi \theta}= \langle \psi_k \theta_k^{\dagger} \rangle$,
 $\C_k^{\theta \theta }= \langle \theta_k \theta_k^{\dagger} \rangle $ with $\langle\;\bullet\;\rangle$ denoting ensemble averaging,
 which under the ergodic assumption is equal to the zonal average.
 The covariance of the stochastic excitation
 \begin{equation}
\Q_k~=~
\left(%
\begin{array}{cc}
\Delta_k^{-1} \F_k \F_k^{\dagger} \Delta_k^{-1 \dagger}& 0 \\
 0 & \Delta_{k \la}^{-1} \F_k^{\theta} \F_k^{\theta \dagger}\Delta_{k \la}^{-1 \dagger}\end{array}%
\right)\ ,
\end{equation}
has been normalized so that for each $k$ a unit of energy per unit time is injected by the excitation and therefore the
amplitude of the excitation is controlled by the parameter $\epsilon$.

The S3T equations for the mean flow~(\ref{eq:NL_mbrcl}) in terms of the perturbation covariances are:
\begin{subequations}
\label{eq:Ubrcl}
\begin{align}
\frac{d U }{d t} &= \sum_k~ \frac{k}{2}~{\rm diag} \left [ \Im \( D^2 \C_k^{\psi \psi} + D^2 \C_k^{\theta \theta} \)\right ]
- \nonumber \\
&~~~~~-\frac{{r}}{2} (U - H) -r_m U+\nu D^2 U, \label{eq:mean1}\\
 \frac{d D_\lambda^2 H }{d t} &= D^2 \sum_k \frac{k}{2}~{\rm diag} \left [ \Im \left (
 D^2_\lambda \C_k^{\psi \theta \dagger} + D^2 \C_k^{\psi \theta} \right ) \right ] + \nonumber \\
 &~~~~~+ \frac{r}{2} D^2(U-H)+2 \lambda^2 r_T \left (
H -\frac{H_T}{2} \right ) -\nonumber\\
 &~~~~~ -r_m D^2_\lambda H + \nu D^2 D^2 H\ . \label{eq:mean2}
 \end{align}
\end{subequations}

As is the case in the previous barotropic examples, for homogeneous forcing there also exists a meridionally and zonally homogeneous
turbulent S3T equilibrium state consisting of an equilibrium zonal mean flow equal to the thermal wind balanced radiative equilibrium flow
${\cal U}^e= [H_T/2, H_T/2]$ corresponding to layer velocities $U_1^e=H_T$ and $U_2^e=0$ and a perturbation field with
covariances, $\C_k^e$ which satisfy the corresponding steady state Lyapunov equations
(\ref{eq:ensemblea}) (for the explicit expression of the equilibrium covariance see \cite{DelSole-Farrell-95}).
However, unlike in the previous case of the barotropic beta-plane example, this homogeneous equilibrium state is realizable only for parameter values
for which ${\cal U}^e$ is hydrodynamically (baroclinically) stable. In the absence of dissipation
instability occurs for this constant flow when  $H_T > \beta / \lambda^2$ or,
in terms of the criticality parameter $\xi = \max(U_1-U_2) \lambda^2/\beta$, when
$\xi >1$. This dissipationless criticality parameter is customarily used despite the fact that $\xi>1$ implies instability only when the
meridionally uniform flow is disipationless. In the presence of dissipation the homogeneous S3T equilibrium ${\cal U}^e= [H_T/2, H_T/2]$
is realizable when $\xi < \xi_c$, where $\xi_c$ is the critical value for instability for the parameters of the flow being studied.
Although with $\xi<\xi_c$ the homogeneous equilibrium state is stable to the traditional baroclinic modal instability,
it becomes S3T unstable at some critical forcing amplitude
$\epsilon_c$. For $\epsilon>\epsilon_c$ the turbulent flow transitions
to an inhomogeneous S3T turbulent fixed point state consisting of zonal jets and their associated perturbation field.
The formation of finite amplitude equilibrium jets under homogeneous forcing in this baroclinically stable
regime occurs through a bifurcation similar to that discussed previously in the example of barotropic jet formation.
When $\xi >\xi_c$ the only homogeneous states are the (unstable) laminar equilibria with $\epsilon=0$ and $\C_k^e=0$
and the flow if perturbed transitions to an inhomogeneous state, which equilibrates to an inhomogeneous
statistical state with zonal jet flows\footnote{In this regime the S3T self-sustains in the absence of forcing. In the absence of forcing the S3T reduces to the corresponding QL simulation.
NL self-sustains when $\xi > \xi_c$, but we note that self-sustaining turbulent states have been found for subcritical parameters close to criticality \citep{Lee-Held-1991}.}.

We wish to examine the equilibria that are predicted by S3T both in the baroclinically stable and unstable regimes
and compare them with the turbulent states in
corresponding QL and NL simulations. We consider  two example cases, one with $\xi=0$, a stable case with no temperature
difference across the channel, and another with $\xi=2$ in which case the temperature gradient is being relaxed to a baroclinically unstable shear.

\begin{figure}[t]
\centering
\includegraphics[width=19pc]{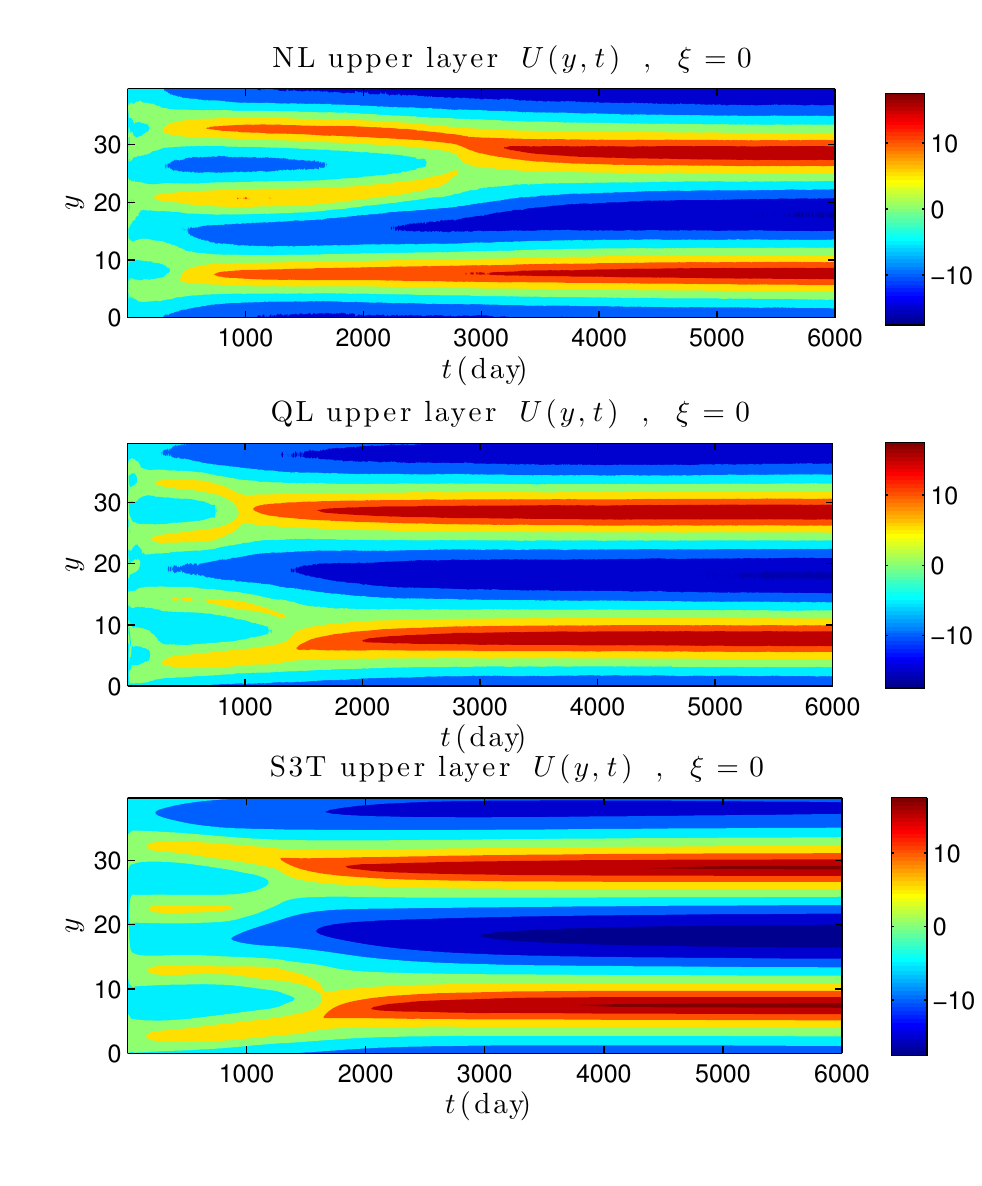}
\caption{Hovm\"{o}ller diagrams of jet emergence in NL, QL and S3T simulations
under Jovian conditions with $\xi=0$. Shown for the NL, QL and S3T simulations are
the flow, $U_1(y, t)$, in the upper layer (the flow in the lower layer is not shown as the flow is almost barotropic).
After a series of mergers S3T is attracted
to a statistical steady state with a $n=2$ jet. The whole process of jet mergers and S3T equilibration is accurately
reflected in the QL and NL simulations. This figure shows
that S3T predicts the structure, growth and equilibration of strongly forced jets in
both the QL and NL simulations. The dissipation parameters in all simulations are
$r=r_T=0$, $r_p=1/5$ and $r_m=1/2000$. The channel has $L_x=L_y=40$ and global zonal wavenumber $k=1,\dots,14$ are
equally excited in energy. The total energy input by the stochastic excitation is $2.5~\textrm{W}\,\textrm{m}^{-2}$.}
\label{fig:hov_jovl}
\end{figure}

\begin{figure}
\centering
\includegraphics[width=19pc]{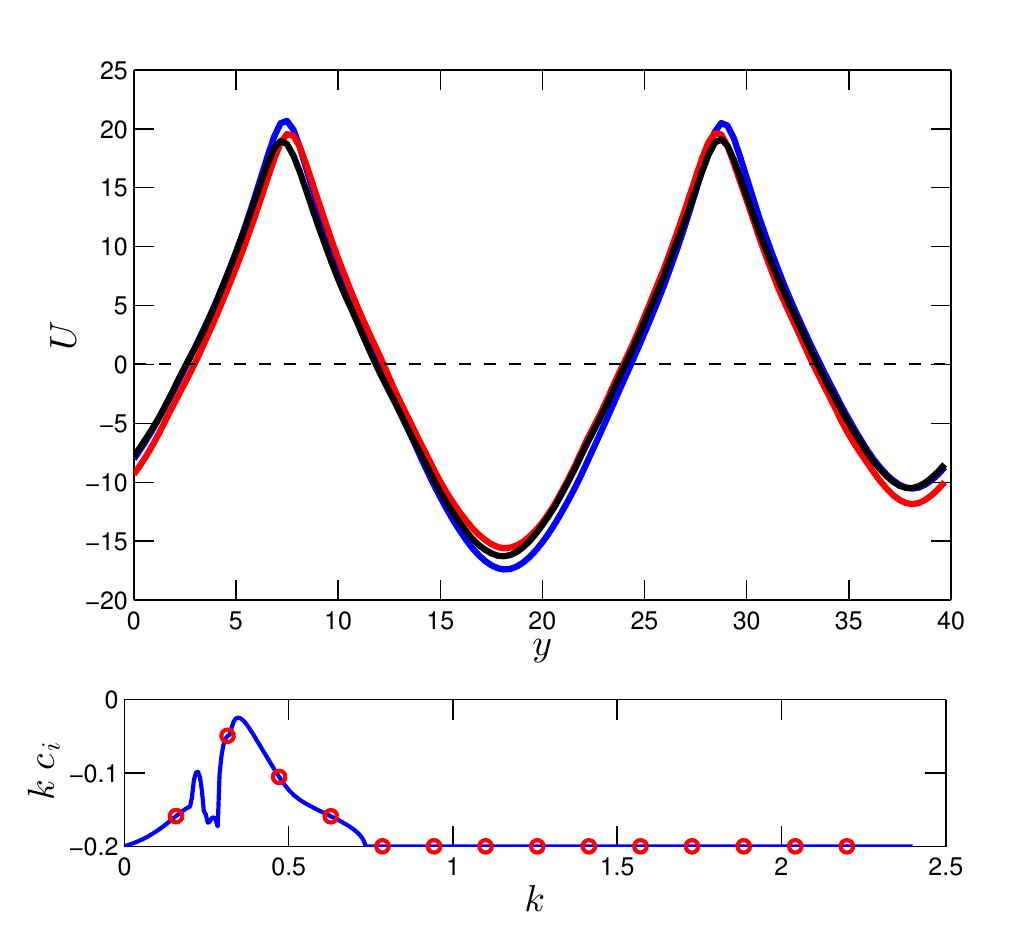}
\caption{Comparison of the S3T equilibrium jet in the upper layer for $\xi=0$ (solid blue) with instantaneous realizations of the jet
in corresponding NL (solid black) and QL (solid red) simulations. The jet is barotropic and the jet in the lower layer is not shown.
The growth rate of the least damped eigenmode of the S3T equilibrium jet is shown as function of zonal wavenumber
in the lower panel. The circles in this plot
indicate the growth rate for each of the $14$ harmonics
retained in the S3T simulation. The parameters are as in Fig.~\ref{fig:hov_jovl}.}
\label{fig:U_jovl}
\end{figure}

When $\xi=0$ the S3T dynamics of the two layer model is found to be essentially barotropic, similar to the examples
in section 1.6, with the
difference being
that the deformation radius is finite. This case models the atmospheres of the
outer planets which have small temperature gradients and are maintained turbulent by injection of energy by small scale convective excitation.
The same stochastic
excitation is applied to the S3T and QL and NL simulations by forcing the 14 gravest zonal wavenumbers, $k$, with the forcing structure
 $\F_k $ chosen so that the $(i,j)$ element of the excitation is
 proportional to $\exp[-(y_i-y_j)^2/(2s^2)]$, as in~(\ref{eq:forcing}),
with $s = 1/\sqrt{2}$. The total energy input by this excitation is $2.5~\rm W m^{-2} $ distributed equally among the excited zonal components. The domain is
a doubly periodic channel with $L_x=L_y=40$. Dissipation parameters for the NL, QL and S3T simulations are $r=r_T=0$, $r_p=1/5$ and $r_m = 1/2000$.
An example of emergence of zonal flows in this system can be seen in Fig. \ref{fig:hov_jovl} in NL, QL and S3T and a comparison of the corresponding zonal mean flows
is shown in Fig.~\ref{fig:U_jovl}.  S3T predicts that for these parameters the statistics of the turbulent flow are attracted to
an equilibrium with a jet and associated consistent eddy structure.
The equilibrium mean flow is barotropic and barotropically stable with maximum growth rates plotted as a function
of $k$ in Fig.~\ref{fig:U_jovl}.  Both the NL and QL simulations reflect the predictions of S3T and the equilibrated flow has the
characteristic complex structure of
the $23^\circ$ N jet on Jupiter with the sharp prograde jets and the rounded retrograde jets \citep{Sanchez-etal-2008, Farrell-Ioannou-2008-baroclinic}.
The structure of the retrograde jets is not quite consistent with
potential vorticity homogenization
\citep{Dritschel-McIntyre-2008}, as the potential vorticity gradient of the equilibrium flow in Fig.~\ref{fig:U_jovl}
is slightly negative in limited locations of the retrograde jets which is indicative of dynamical
processes beyond mixing  (cf. also Fig.~3 in \cite{Farrell-Ioannou-2008-baroclinic}).

 \begin{figure}[t]
\centering
\includegraphics[width=19pc]{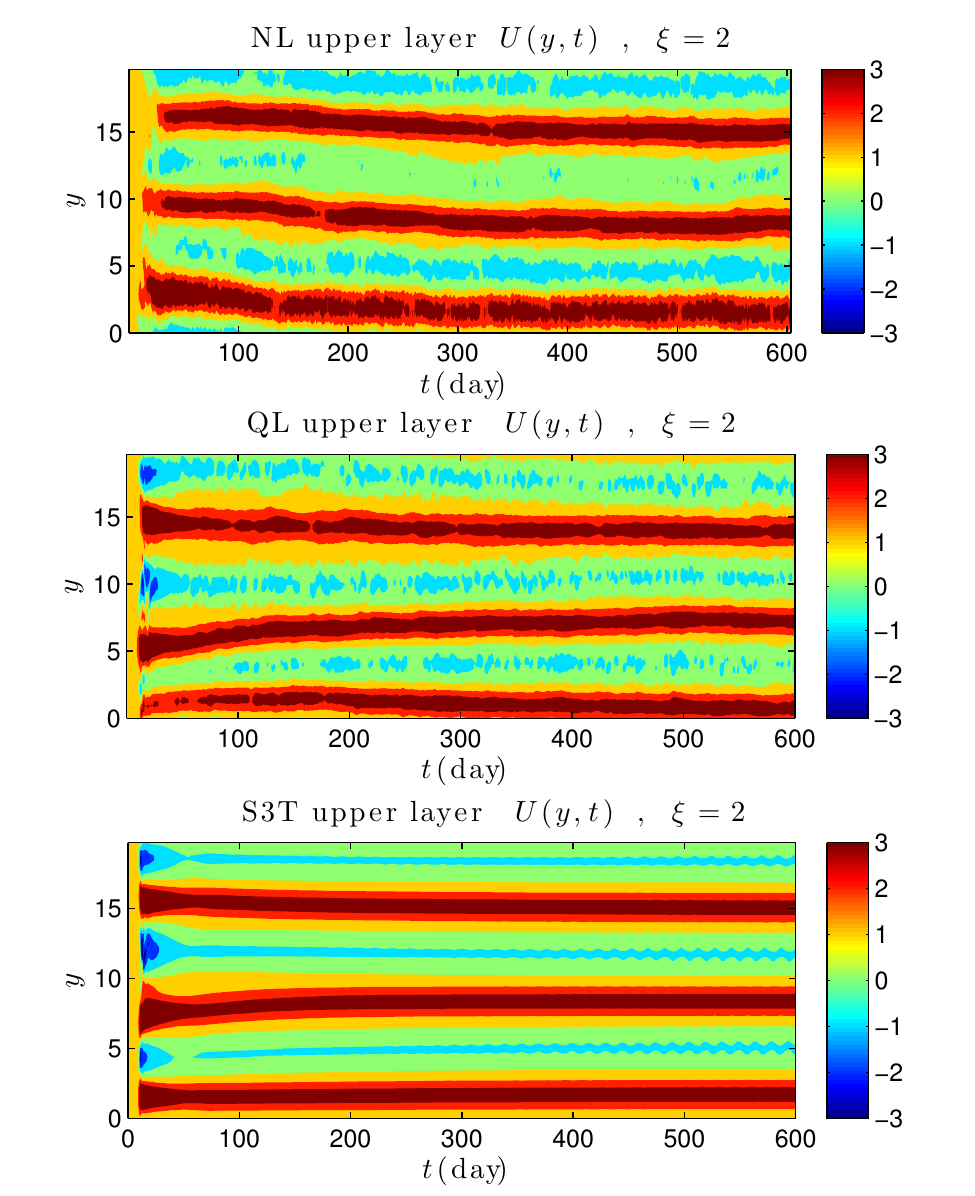}
\caption{Hovm\"{o}ller diagrams of jet emergence in NL, QL and S3T simulations
under Earth-like conditions with supercritical $\xi=2$. Shown for the NL, QL and S3T simulations are
the flow, $U_1(y, t)$, in the upper layer. The flow has substantial baroclinicity
and mean flow in both layers is shown in Fig.~\ref{fig:U_xi2}. This figure shows
that S3T predicts the structure, growth and equilibration of strongly forced jets in
both the QL and NL simulations of self-sustained baroclinic turbulence.
The dissipation parameters are
$r=1/10$, $r_T=1/20$, $r_p=1/5$ and $r_p=1/5$ and
the channel size is $L_x=40$ and $L_y=20$. In the QL and S3T simulations
global zonal wavenumber $k=1,\dots,14$ are stochastically excited equally in energy with a total
energy injection of $0.5~\textrm{W}\,\textrm{m}^{-2}$. In QL and S3T diffusive damping is included with $\nu=0.02$.}
\label{fig:brclhov_xi2}
\end{figure}

\begin{figure}
\centering
\includegraphics[width=17pc]{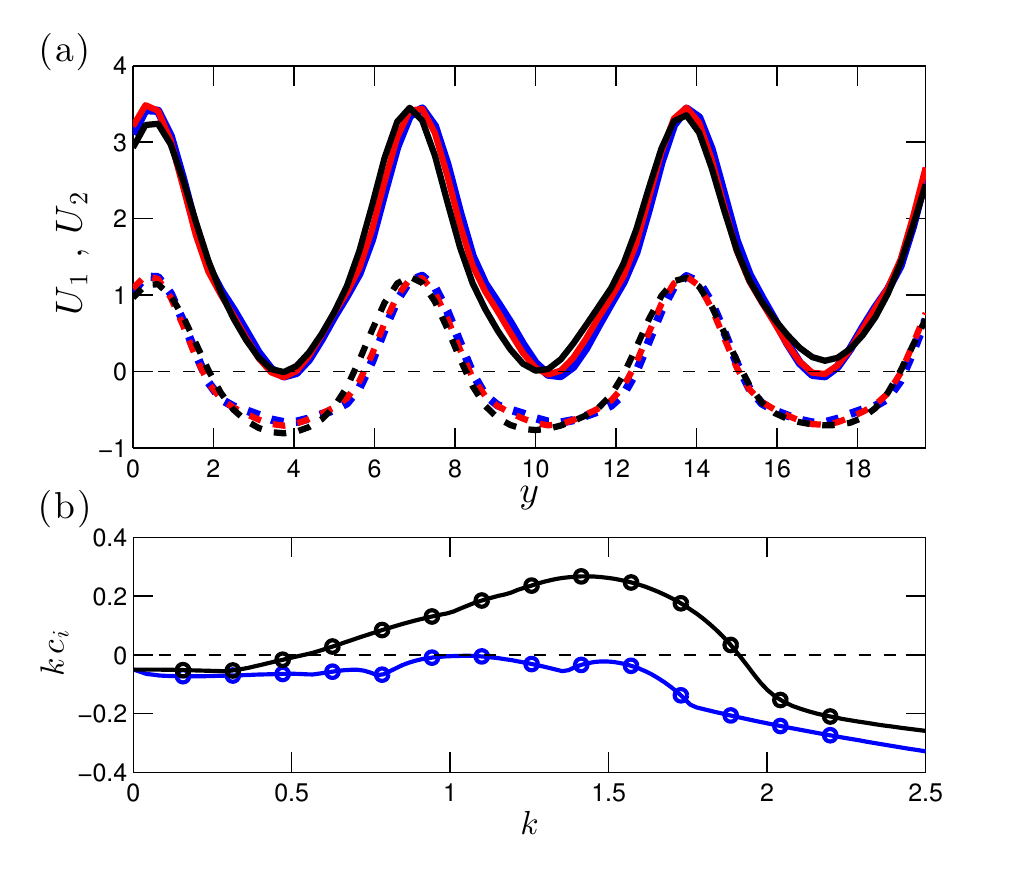}
\caption{
(a) Upper and lower layer flow in S3T (blue), QL (red) and NL (black) for supercritical equilibria with $\xi=2$.
(b) Growth rate as a function of zonal wavenumber $k$ (blue curve) for
the S3T equilibrated flow shown in (a) and in black the corresponding growth rate of
the meridionally uniform baroclinic flow with $\xi=2$  with $r_T=1/15$ and $r=1/5$.
With these dissipation parameters the flow becomes unstable for $\xi > \xi_c = 0.875$. In the absence of
dissipation the shortwave cut-off occurs at $k=\lambda^2=2$. The circles
indicate the growth rate of the $k$ associated with the $14$ harmonics
used in the S3T calculations. The parameters are as in Fig.~\ref{fig:brclhov_xi2}.}
\label{fig:U_xi2}
\end{figure}

Consider now the baroclinically unstable case, $\xi=2$, which represents midlatitude Earth like conditions. We again
choose
a doubly periodic channel with
$L_x=40$ and $L_y=20$ and impose dissipation parameters
$r=1/10$, $r_T=1/20$ and $r_p=r_m=1/5$.
For these dissipation parameters the shear $H_T$ is baroclinically unstable
(cf. a plot of the maximum growth rate of this flow as a function of zonal wavenumber is shown in Fig.~\ref{fig:U_xi2}c) and
a self-sustained turbulent state develops in NL, QL and S3T. In the absence of excitation
the NL simulation equilibrates to a 3 jet structure as
shown in the top panel of
Fig.~\ref{fig:brclhov_xi2} with associated jet structure
shown in  Fig.~\ref{fig:U_xi2}a.
To obtain correspondence with the NL we parameterize the
eddy interactions in
the QL and S3T as in \cite{DelSole-Farrell-1996},
using a state-independent stochastic forcing, with the same structure as that
used for $\xi=0$. A total injection rate
of $0.5~\rm W m^{-2}$ in the 14 gravest zonal components and diffusion with
$\nu=0.02$ bring the S3T and the QL simulations in good agreement with the
NL. The agreement in the emergence of the jets and in the maintained flows are shown in
Fig.~\ref{fig:brclhov_xi2} and Fig.~\ref{fig:U_xi2}. S3T predicts that for these parameters the statistics of the turbulent flow are attracted to
an equilibrium with a jet and associated eddy structure. The mean flow is by necessity
stable with growth rates shown in Fig.~\ref{fig:U_xi2}b. To avoid instability
the jets become increasingly east/west asymmetric as criticality is increased with
the eastward-portion equilibrating by zonal confinement \citep{Ioannou-Lindzen-1986, James-1987, Roe-Lindzen-1996},
and the westward jets equilibrating more barotropically approximately close to the Rayleigh-Kuo stability boundary
while maintaining substantial baroclinicity.

\begin{figure}
\centering
\includegraphics[width=19pc,trim = 12mm 0mm 12mm 0mm, clip]{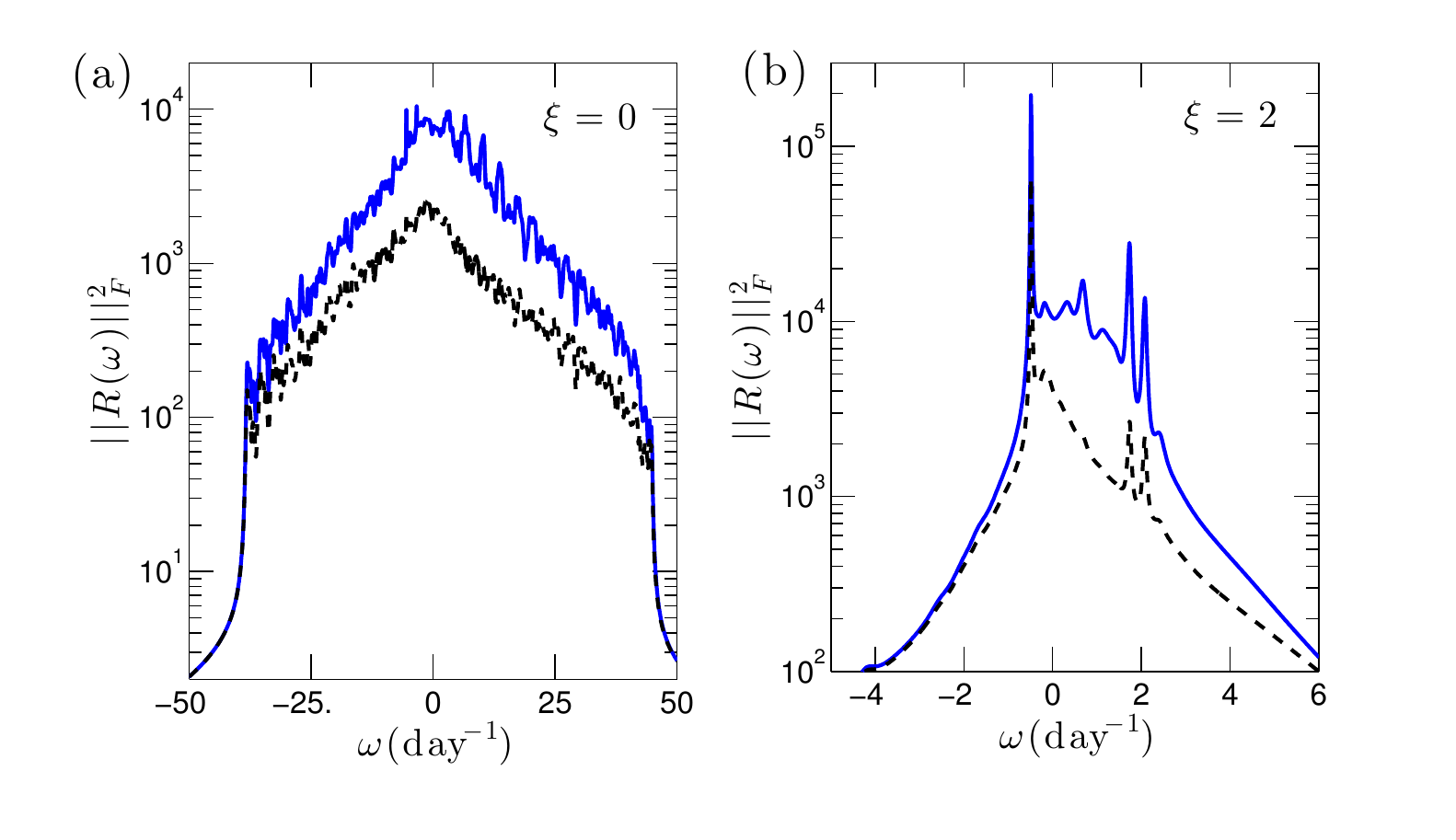}
\caption{
Panel (a): the
Frobenius norm of the resolvent associated with the eddy dynamics about the equilibrium mean flow of Fig.~\ref{fig:U_jovl}
 that obtains at $\xi=0$ as a function of frequency and the Frobenius norm of the resolvent
of the corresponding equivalent normal eddy dynamics (dashed).
Similarly in (b) for the equilibrium flow shown in Fig.~\ref{fig:U_xi2} for $\xi=2$. }
\label{fig:1}
\end{figure}

A characteristic of the equilibrated jets in all cases is that
while stable, they are highly non-normal and support strong transient growth.
This non-normal growth is succinctly summarized
in Fig.~\ref{fig:1} (a,b)
by comparison between the
Frobenius norm of the resolvent of the jet perturbation dynamics
\begin{equation}
\left\| \R (\omega) \vphantom{R^k_k} \right\|^2_F = \sum_k {\rm trace} \left ( \R_k(\omega) \R_k(\omega)^{\dagger} \right )\ ,
\end{equation}
where
\begin{equation}
\R_k(\omega) =- \left(\i \omega \I +  \A_k(U^e) \right)^{-1}\ ,
\end{equation}
and its equivalent normal counterpart, with
resolvent the diagonal matrix, $\S_k$, of the eigenvalues of $\A_k$:
\begin{equation}
{\R}^\perp_{k}(\omega) =- \left(\i \omega \I + \S_k\right)^{-1}\ .
\end{equation}
The square Frobenius norm of the resolvent shown  as a function of frequency, $\omega$, in Fig.~\ref{fig:1}
is the ensemble mean eddy streamfunction variance, $\langle | \psi |^2 +| \theta |^2 \rangle $, that would be maintained
by white noise forcing of this equilibrium jet.
Non-normality increases the maintained streamfunction variance $\langle | \psi |^2 +| \theta |^2 \rangle$,
over that of the equivalent normal system $\langle |\psi^{\perp}|^2 + |\theta^{\perp}|^2\rangle $ and the extent of this increase is a measure of the
non-normality \citep{Farrell-Ioannou-1994b,Ioannou-1995}. The non-normality of the equilibria, which is associated with both baroclinic and barotropic growth processes,
increases as $\xi$ increases as shown in Fig.~\ref{fig:1}.

\begin{figure}
\centering
\includegraphics[width=16pc]{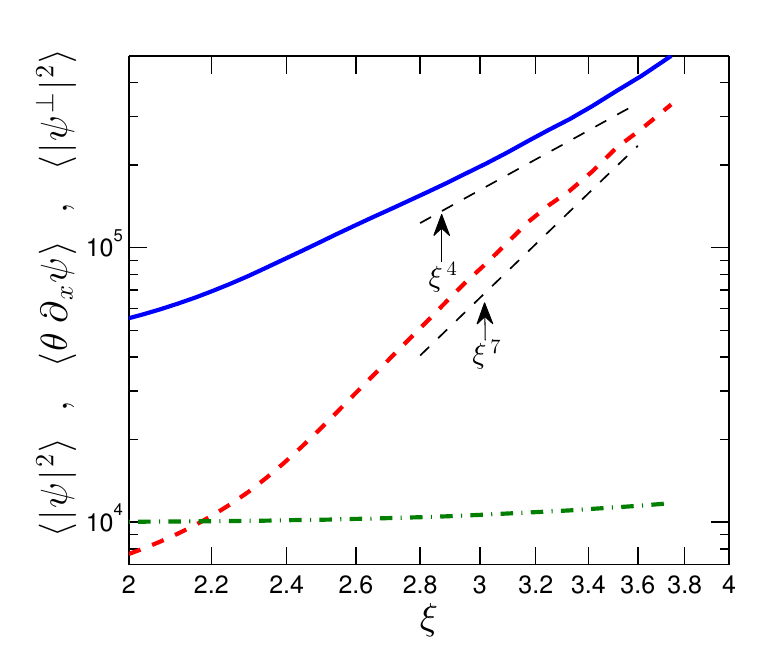}
\caption{For the equilibrated jets shown are: the eddy barotropic streamfunction variance $\langle | \psi |^2 \rangle$ (solid line),
the eddy heat flux $\langle \theta \partial_x \psi \rangle$ (dashed line),
 and the eddy barotropic streamfunction
 variance maintained by the equivalent normal system $\langle | \psi^\perp | ^2 \rangle $ (dash-dot line)  as a function of the criticality parameter $\xi$.
The eddy  barotropic streamfunction variance increases
 as $\xi^4$, the heat flux increases as $\xi^7$, while the equivalent normal variance is nearly constant.
 The dissipation parameters are $r=1/10$, $r_T=1/20$, $r_p=1/5$, $r_m=1/5$, $\nu=0$ and
the channel size is $L_x=40$ and $L_y=20$. The 14 gravest zonal components are excited.}
\label{fig:2}
\end{figure}

The maintained eddy streamfunction variance, $\langle | \psi |^2 \rangle$,  in the equilibrium jet and for comparison
the eddy streamfunction variance, of the equivalent normal jet system,
 $\langle |\psi^{\perp}|^2 \rangle$,
 as a function of the criticality measure of the equilibrated flow $\xi$ is shown Fig.~\ref{fig:2}.
The eddy variance maintained
by the equilibrium jet increases as $\xi^4$ while the equivalent normal system eddy
variance, $\langle |\psi^{\perp}|^2 \rangle$, does not
increase appreciably.
This increase in variance
with criticality is due to the increase in the non-normality of the equilibrated jet.
The heat flux, which is proportional to
$\langle { \theta \partial_x \psi} \rangle$,
exhibits a $\xi^7$ power law behavior implying an equivalent higher order thermal diffusion.
While such power law behavior is recognized to be generic to strongly turbulent equilibria
\citep{Held-Larichev-1996, Barry-etal-2002, Zurita-Gotor-2007}, it lacked comprehensive explanation in the absence of the
SSD-based theory for the structure of the statistical equilibrium turbulent state and its concomitant
non-normality that is provided by S3T. Coexistence of very high non-normality with modal stability has been heretofore regarded as highly unlikely to occur naturally and such systems have been
generally thought to result from engineering contrivance. In fact the goal of much of classical control theory
is to suppress modal instability by designing stabilizing feedbacks.
That this process of modal stability suppression sometimes resulted in modally stable systems
that were at the same time highly vulnerable to disruption due to large transient growth of perturbation
led to the more recent development of robust control theory which seeks to control transient growth associated
with non-normality as well as modal stability \citep{Doyle_Book}.
A widely accepted argument for the necessity of engineering intervention to produce a system
that is at the same time modally stable and highly non-normal follows from the observation that
in the limit of increasing system non-normality, arbitrarily small perturbations to the dynamics result
in modal destabilization. If the dynamics is expressed using a dynamical matrix, as we have done
here, this vulnerability of the non-normal dynamics to modal destabilization finds expression in the
pseudo-spectrum of the dynamical matrix \citep{Trefethen-2005}. It is remarkable that the naturally
occurring feedback between the zonal mean flow and the perturbations in turbulent systems stabilizes these systems.

In summary, we have provided an explanation for the observation that
adjustment to stable but highly amplifying states exhibiting
power law behavior for flux/gradient relations is characteristic of strongly supercritical baroclinic turbulence.
One consequence of stability coexisting with  high non-normality in the Earth's midlatitude atmosphere is the association of
cyclone formation with chance occurrence in the turbulence of optimal or near optimal
initial conditions \citep{Farrell-1982, Farrell-1989, Farrell-Ioannou-1993d, DelSole-2004b}.
We have explained why a  state
of high non-normality together with marginal stability is inherent: it is
because the SSD of the turbulence maintains flow stability by adjusting the system to be in
the vicinity of a specific stability boundary,
which is identified with the fixed point of the S3T equilibrium,
while retaining the high degree of non-normality of the system.
Such a state of extreme non-normality coexisting with exponential stability is an emergent consequence of the underlying SSD of
baroclinic turbulence
and can only result
from the feedback control mechanism operating in baroclinic turbulence which has been identified by SSD analysis using S3T.

\section{Application of S3T to study the SSD of wall-bounded shear flow turbulence}

Consider plane Couette flow between walls with velocities $\pm U_w $.
 The streamwise direction is $x$, the wall-normal direction is $y$, and the spanwise direction is $z$.
 Lengths are non-dimensionalized by the channel half-width, $\delta$, and
velocities by $U_w$, so that the Reynolds number is $R= U_w \delta / \nu$, with $\nu$ the coefficient of kinematic viscosity.
We take for our example a doubly periodic channel of non-dimensional length $L_x$ in the streamwise direction and $L_z$ in the spanwise
(note we have adopted the wall-bounded turbulence convention for the vertical coordinate which differs from the meteorological convention).

In the absence of turbulence the equilibrium solution is the laminar Couette flow with velocity components $(y,0,0)$
which has been shown to be linearly stable at all Reynolds numbers \citep{Romanov-73}
and globally stable for $R<20.7$ \citep{Joseph-1966}. However, experiments show
that plane Couette flow
can be induced by a perturbation to transition to a turbulent state for Reynolds numbers exceeding
$R \approx 360$ \citep{Tillmark-1992}.  During transition to turbulence and in the turbulent state a
prominent
large scale structure is observed in the flow. This structure, referred to as the roll/streak, comprises
 a modulation of the streamwise mean flow in the spanwise
direction by regions of high and low velocity, referred to as streaks, together with a set
of nearly cylindrical vortices in the wall-normal/spanwise plane,
referred to as rolls. The roll circulation is such that the maximum negative wall-normal velocity is
coincident with the maximum positive streamwise velocity of the streak and the maximum wall-normal velocity with the minimum streak velocity.
This roll circulation serves to amplify the streak by advecting the mean shear; a mechanism
referred to as lift-up.
The streak  is analogous to the jet that develops in barotropic and baroclinic flows and
roll-streak structure arises naturally
from S3T instability of the spanwise homogeneous shear turbulence. To show this
we formulate the S3T dynamics for this flow and demonstrate that the S3T
homogeneous turbulent equilibria become unstable with the eigenfunctions of maximum growth rate being
the roll-streak structures.

Consider the vector velocity field $\vec{ \it U}$ to be decomposed into a streamwise mean, with components, $(U,V,W)$, and
perturbation from this mean with components $(u,v,w)$.
The pressure gradient is similarly decomposed into its streamwise mean, $\nabla P$, and perturbation from this mean, $\nabla p$.
All streamwise averaged quantities are denoted with capitals and the streamwise averaging operation is
denoted by an overline. In these variables a
unit density fluid obeys the non-divergent Navier-Stokes equations:\begin{subequations}
\label{eq:NSE0}
\begin{align}
&{ \vec{\it u}_t}+  \vec{ \it U} \cdot \nabla \vec{\it u} +
\vec{\it u} \cdot \nabla \vec{\it U} + \nabla {p} - \Delta { \vec{\it u}} / R = \nonumber \\
& \hspace{2em}=- ( \vec{\it u} \cdot \nabla \vec{\it u} -
 \overline{\vec{\it u} \cdot \nabla \vec{\it u}} ) +\vec{\it e}\ ,
\label{eq:NSp}\\
&{ \vec{\it U}_t}+  \vec{\it U} \cdot \nabla \vec{\it U}
 + \nabla {P} - \Delta { \vec{\it U}} / R = -
 \overline{\vec{\it u} \cdot \nabla \vec{\it u}} \ ,
\label{eq:NSm} \\
& \nabla \cdot \vec{\it U} = 0 ~\ ,~~\nabla \cdot \vec{\it u} = 0\ ,\label{eq:ND}
\end{align}\end{subequations}with
boundary conditions $\vec {\it u} =0$ and $\vec{\it U} = (\pm 1, 0, 0)$ at $y=\pm 1$ and periodicity in $x$ and $z$.

In the perturbation equation~(\ref{eq:NSp}) allowance is made for specifying
an explicit external perturbation forcing, $\vec{\it e}$.
A stochastic parameterization is now introduced to account for both this external forcing and the
perturbation-perturbation interactions,  $\vec{\it u} \cdot \nabla \vec{\it u} -\overline{\vec{\it u} \cdot \nabla \vec{\it u}}$.
With this parameterization,  the perturbation equation becomes:
\begin{equation}
{ \vec{\it u}_t}+  \vec{\it U} \cdot \nabla \vec{\it u} +
\vec{\it u} \cdot \nabla \vec{\it U} + \nabla {p} - \Delta { \vec{\it u}} / R = \vec{\it E}\ . \label{eq:NSpf}
\end{equation}
Perturbation equation~(\ref{eq:NSpf}), coupled with the mean flow equation,
(\ref{eq:NSm}), form the quasi-linear (QL) form of the Navier-Stokes equations.
This approximate set of equations is also referred to as the restricted nonlinear (RNL) system \citep{Thomas-etal-2014}.

\begin{figure}
\centering
\includegraphics[width=17pc]{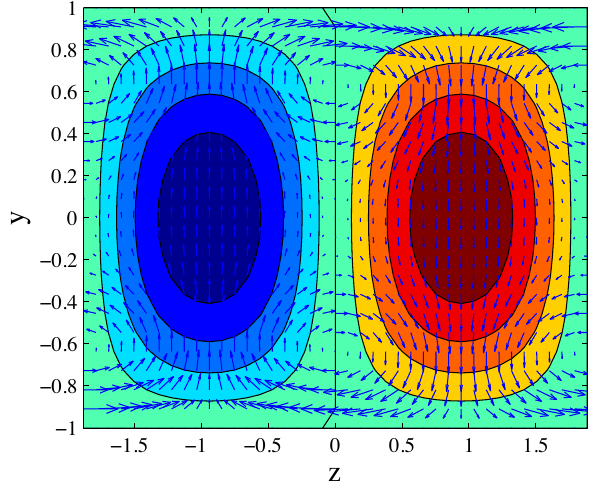}
\caption{ The rate of change of streamwise roll acceleration induced by a streak perturbation to a
Couette flow that is maintained turbulent by stochastic forcing.  Distortion
of the turbulence by the streak perturbation induces Reynolds stresses that force roll circulations supporting the streak via the lift-up mechanism.  Shown are
contours of  the imposed streak perturbations, $ \delta  U =\cos(\pi y/2) \sin(2 \pi z / L_z)$,  with $\delta U>0$ in $z>0$,  and vectors of the resulting rate
of change of roll acceleration, $(\dot V, \dot W)$.
The Reynolds number is $R=400$, $L_x=1.75 \pi$ and $L_z=1.2 \pi$. Adapted from 
``Dynamics of streamwise rolls and streaks in turbulent wall-bounded shear flow'', by B. F. Farrell and P. J. Ioannou, 2012, Journal of Fluid Mechanics, vol. 708, pp. 149-196. \textcopyright\ Cambridge University
Press. Reprinted with permission.}
\label{fig:benney_random}
\end{figure}

It is convenient to eliminate the pressure and express equation~(\ref{eq:NSpf}) in terms of cross-stream velocity $v$ and  cross-stream vorticity,
$\eta=\partial_z u -\partial_x w$. The equations then take the form:
\begin{subequations}
\label{eq:PNSE}
\begin{align}
&\Delta v_t +  U \Delta v_x + U_{zz} v_x + 2 U_z v_{xz} - U_{yy} v_x - 2 U_z w_{xy}-
\nonumber \\
& \hspace{2em}- 2 U_{yz} w_x - \Delta \Delta v/R = \Delta E_v\ , \label{eq:pv} \\
&\eta_t + U \eta_x - U_z v_y + U_{yz} v + U_y v_z+ U_{zz} w  - \Delta \eta/R  = E_{\eta}\ .
\label{eq:peta}
\end{align}
\end{subequations}
 where $E_v$ and $E_{\eta}$
 are the stochastic excitation in these variables (cf. \cite{Schmid-Henningson-2001}).
In the perturbation equations~(\ref{eq:PNSE}), advection of perturbations by
the small $V$ and $W$ components of the streamwise mean velocity
has  been neglected\footnote{The results presented are not affected by
neglecting the advection of the perturbation field by $V$ and $W$ velocities in the perturbation equations, cf.~\cite{Thomas-etal-2014}.}.
Using nondivergence the mean flow equation~(\ref{eq:NSm}) can be written as:
\begin{subequations}
\label{eq:MNSE}
\begin{align}
U_t &= U_y \Psi_z - U_z \Psi_y -\partial_y \overline{uv} -\partial_z \overline{uw} + \Delta_1 U/R\ ,\label{eq:MU}\\
\Delta_1 \Psi_t& = (\partial_{yy}-\partial_{zz})(\Psi_y \Psi_z - \overline{ vw} ) - \nonumber \\
& \hspace{2em}- \partial_{yz}( \Psi_y^2 - \Psi_z^2 + \overline{w^2} - \overline{v^2} ) + \Delta_1 \Delta_1 \Psi /R\ .\nonumber \\
\label{eq:MPSI}
\end{align}
\end{subequations}
In~(\ref{eq:MPSI}), $\Delta_1\equiv \partial_{y}^2+\partial_{zz}^2$ and $V$ and $W$ have been
expressed in terms of the streamfunction, $\Psi$, as $V= - \Psi_z$ and $W = \Psi_y$.

 \begin{figure}
\centering
\includegraphics[width=19pc,trim = 9mm 0mm 5mm 0mm, clip]{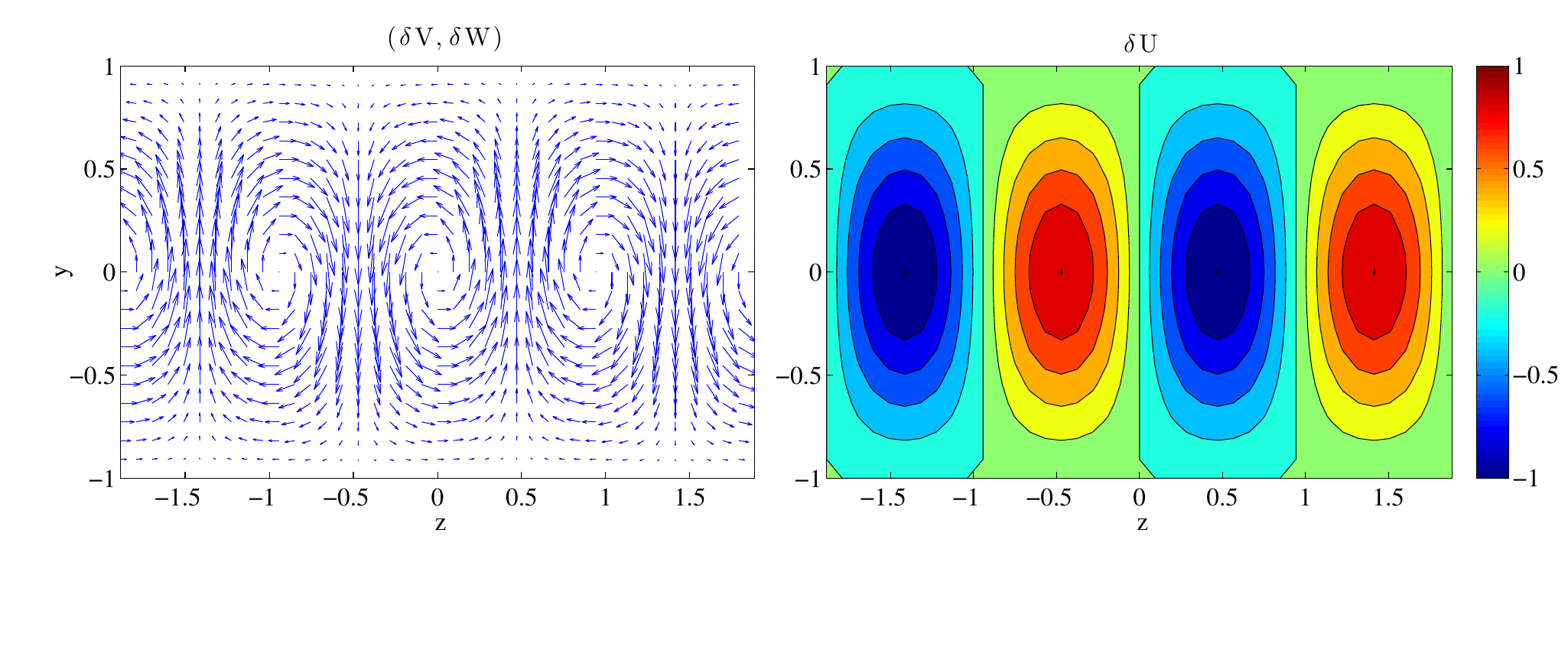}
 \caption{The most unstable streamwise roll and streak eigenfunction of the S3T system
 linearized about the spanwise uniform equilibrium at supercriticality $\epsilon/\epsilon_c=1.4$.
 The growth rate of this mode is $\lambda_r = 0.014$. Shown are  velocity vectors $(\delta V, \delta W)$ (left) and streamwise velocity $\delta U$ (right). The ratio of the maxima of
$(\delta U, \delta V, \delta W)$ is $(1,0.06,0.03)$.
Other parameters are as in Fig.~\ref{fig:benney_random}. Adapted from 
``Dynamics of streamwise rolls and streaks in turbulent wall-bounded shear flow'', by B. F. Farrell and P. J. Ioannou, 2012, 
Journal of Fluid Mechanics, vol. 708, pp. 149-196. \textcopyright\ Cambridge University
Press. Reprinted with permission.}
\label{fig:8}
\end{figure}

 We next Fourier expand the perturbation fields in $x$: $v= \Re \left[ \sum_k \hat v_{k}(y, z, t) {e}^{\i k x}\right]$, $\eta= \Re\left[\sum_k \hat \eta_{k}(y, z, t) {e}^{ \i k x}\right]$, and write the equations for the evolution of the Fourier components of~(\ref{eq:PNSE}) in the matrix form
\begin{equation}
\frac{ d \phi_k }{ d t}
= \A_k({ U}) \phi_k + \sqrt{\epsilon} \, \F_{k} dB_{tk }  \ ,\
\label{eq:Aphi}
\end{equation}
where the state of the system $\phi_k = [\hat{v}_k,\hat{\eta}_k]^T$ comprises the values of the $\hat{v}_k$ and $\hat{\eta}_k$ on the $N=N_y N_z$
grid points of the $(y,z)$ plane and
\begin{align}
\A_k({ U}) &=
\left(
\begin{array}{cc}
 \L_{OS} & \L_{C_1}  \\
 \L_{C_2} &    \L_{SQ}
\end{array}
\right)\ ,
\end{align}
with
\begin{subequations}
\label{eq:whole}
\begin{align}
\L_{OS} & = {\Delta}^{-1}
 \left [  -\i k { U} \Delta +\i k ({ U}_{yy}-{ U}_{zz}) - 2 \i k { U}_z \partial_z -  \vphantom{- 2 \i k ({ U}_z \partial_{yyz}^3+ { U}_{yz} \partial_{yz}^2)\Delta_2^{-1}+ {\Delta \Delta}/{R}} \right .\nonumber\\
 &\qquad \left . - 2 \i k ({ U}_z \partial_{yyz}^3+ { U}_{yz} \partial_{yz}^2)\Delta_2^{-1}+ {\Delta \Delta}/{R}  \right ] , \label{subeq:1}\\
\L_{C_1} & = 2 k^2 \Delta^{-1} \left ( { U}_z \partial_y + { U}_{yz} \right ) \Delta_2^{-1} , \label{subeq:2}\\
\L_{C_2} &= { U}_z \partial_y - { U}_y \partial_z -{ U}_{yz} + { U}_{zz} \partial_{yz}^2 \Delta_2^{-1} , \label{subeq:3}\\
 \L_{SQ} & =   -\i k { U} \Delta + \i k { U}_{zz} \Delta_2^{-1} + {\Delta}/{R}~ , \label{subeq:4}
\end{align}
\end{subequations}
being the conventionally designated Orr-Somerfeld, coupling, and Squire operators respectively.
In equations~(\ref{eq:whole}), $\Delta^{-1}$ and $\Delta_2^{-1}$ are the inverses of the
 matrix Laplacians, $\Delta$ and $\Delta_2 = \partial_{x}^2+\partial_{z}^2$, which are rendered invertible by
enforcing the boundary conditions. The boundary conditions satisfied by the Fourier amplitudes of the perturbation fields are:
periodicity in $x$ and $z$ and $\skew2\hat{{v}}_k= \partial_y \skew2\hat{{v}}_k =\skew2\hat{{\eta}}_k=0$ at $y=\pm1$.

\begin{figure}
\centering
\includegraphics[width=19pc,trim = 6mm 0mm 6mm 0mm, clip]{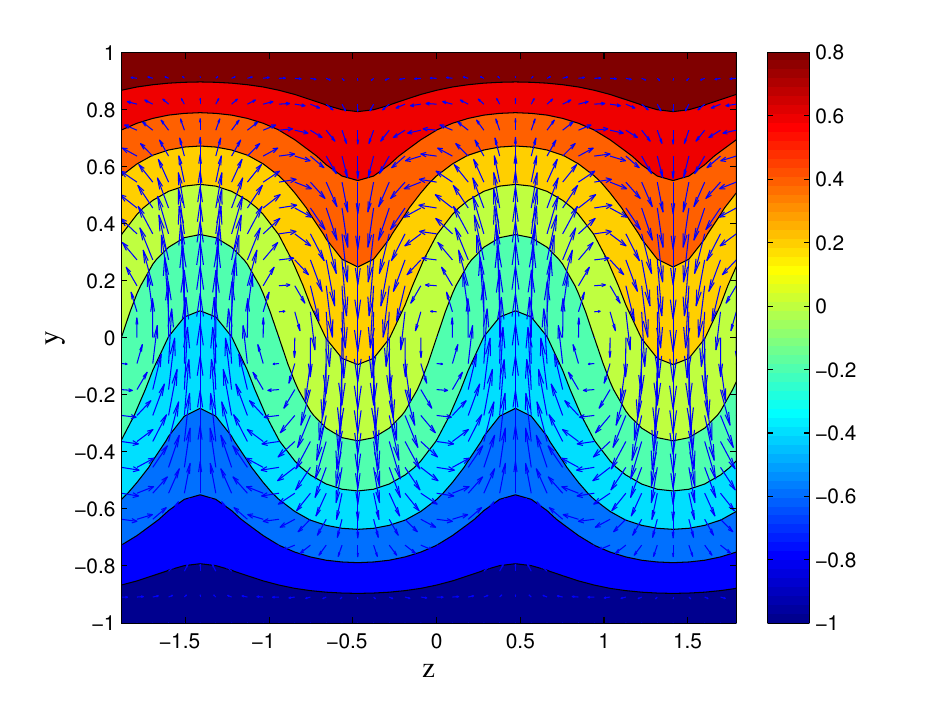}
 \vspace*{-1mm} \caption{
The finite amplitude S3T equilibrium streamwise roll and streak resulting from the equilibration of the
eigenmode  shown in Fig.~\ref{fig:8} at supercriticality $\epsilon/\epsilon_c=1.4$.
Shown are the streamwise averaged streamwise flow, $U(y, z)$, (contours) and the streamwise averaged velocities, $(V, W)$ (vectors). The maxima of the fields $( U, V, W)$ are $(0.26,0.02,0.009)$. Adapted from 
``Dynamics of streamwise rolls and streaks in turbulent wall-bounded shear flow'', by B. F. Farrell and P. J. Ioannou, 2012, 
Journal of Fluid Mechanics, vol. 708, pp. 149-196. \textcopyright\ Cambridge University
Press. Reprinted with permission.}
\label{fig:flow8425}
\end{figure}

The ensemble average perturbation covariance, $\C_k = \langle \phi_k \phi_k^{\dagger} \rangle$,
evolves according to the time-dependent Lyapunov equation:
\begin{equation}
 \frac{{d} \C_k }{{d} t}~ = \A_k( {U} ) \,\C_k +\C_k \,\A_k^{\dagger}({ U} ) + \epsilon\, \Q_k\ ,
  \label{eq:Lyap1}
 \end{equation}
in which: $\Q_k=\F_k \F_k^{\dagger}$. {If then, as previously, we make the ergodic assumption that
streamwise averages are equal to ensemble averages} all the quadratic fluxes that enter
into
the streamwise averaged flow equations~(\ref{eq:MNSE}) become linear functions
of the $\C_k$ and the mean flow evolution equations~(\ref{eq:MNSE})  can be written concisely in the form:
\begin{equation}
\label{eq:MEAN}
\frac{ {d} {\Gamma}}{{d} t} = {G}( {\Gamma}) + \sum_{k} {\Re} \left( \L_{RS} \C_{k} \right )\ ,
\end{equation}
where ${\Gamma} \equiv [{U},{\Psi}]^T$ determines the three components of the streamwise averaged flow,
the term $\sum_{k} {\Re} \left ( \L_{RS} \C_{k} \right )$
produces by multiplying $\C_k$ with matrix $\L_{RS}$ the forcing of the mean equations by the perturbation
field and $G(\Gamma)$ is the nonlinear term representing the self-advection of the streamwise averaged flow.
Equations~(\ref{eq:Lyap1}) and~(\ref{eq:MEAN}) comprise the S3T system for the Couette problem.
The forcing covariances, $\Q_k$, are chosen
 to be spanwise homogeneous. Under this assumption spanwise homogeneous S3T equilibrium states exist.
For further details on the formulation see \cite{Farrell-Ioannou-2012}.

The Couette flow is a laminar equilibrium of the S3T system with excitation $\epsilon=0$.
For any $\epsilon$ and any spanwise homogeneous $Q_k$ there are always
spanwise independent S3T equilibria
having spanwise independent streamwise averaged flow ${U}^e(y)$ and  ${\Psi}^e=0$
and spanwise homogeneous perturbation covariances $\C^e_{k}$.
These are equilibria because,
consistent with the spanwise independence of both the equilibrium mean flow and the imposed excitation,
$\C^e_{k}$ is also spanwise independent and this results in  ensemble mean
$\overline{u w}$, $\overline{v^2}$ and $\overline{w^2}$ that are independent of $z$, and $\overline{v w}$
that identically vanishes by symmetry. Consequently,~(\ref{eq:MPSI}) admits
$\Psi^e=0$ as a solution as the ensemble mean
perturbation forcing vanishes. However, the ensemble mean Reynolds stress divergence
$\partial_y \overline{u v}$ in~(\ref{eq:MU})
does not vanish and ${U}^e(y)$ satisfies $\Delta_1 U^e/R=\partial_y \overline{uv}$ indicating that
the presence of external excitation induces a modification to the laminar Couette profile.

\begin{figure}
\centering
\includegraphics[width=19pc]{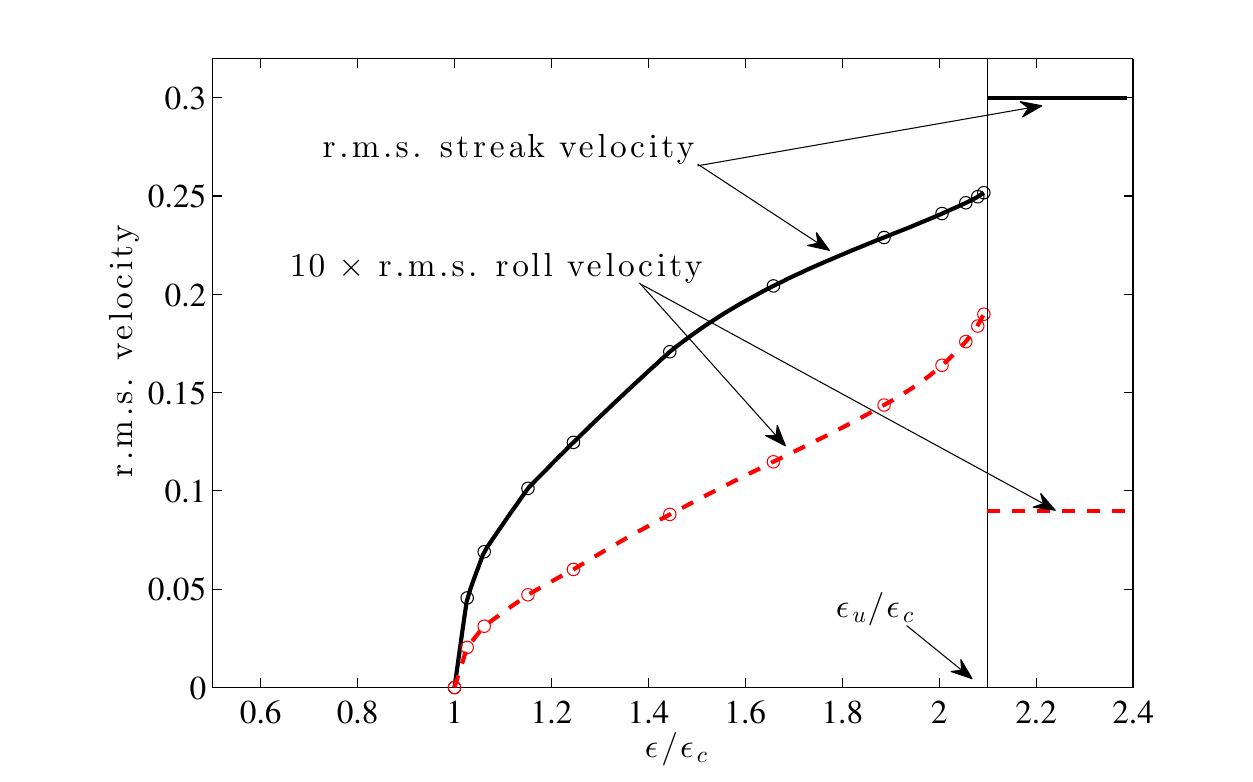}
 \caption{
Typical S3T bifurcation diagram for the Couette problem. Shown are the RMS streak velocity (solid) and $10 \times$ the
RMS streamwise roll velocity (dashed) as a function of the perturbation forcing amplitude, $\epsilon$.
For $\epsilon/\epsilon_c <1 $, the spanwise homogeneous state is S3T stable.
 At $\epsilon_c$ the spanwise uniform equilibrium bifurcates to an equilibrium with a streamwise roll and streak.
Stable streamwise roll and streak equilibria extend up to $\epsilon_u/\epsilon_c=2.1$
beyond which the streamwise roll and streak  transitions to a time-dependent state which can self-sustain
and the amplitudes of the roll and streak become independent of $\epsilon$.
Shown for reference are the r.m.s. velocities of the streak and roll in the self-sustaining state.
The Reynolds number is $R=400$, $L_x=1.75 \pi$ and $L_z=1.2 \pi$. }
\label{fig:HKWbifur}
\end{figure}

\begin{figure}[t]
\centering
\includegraphics[width=19pc]{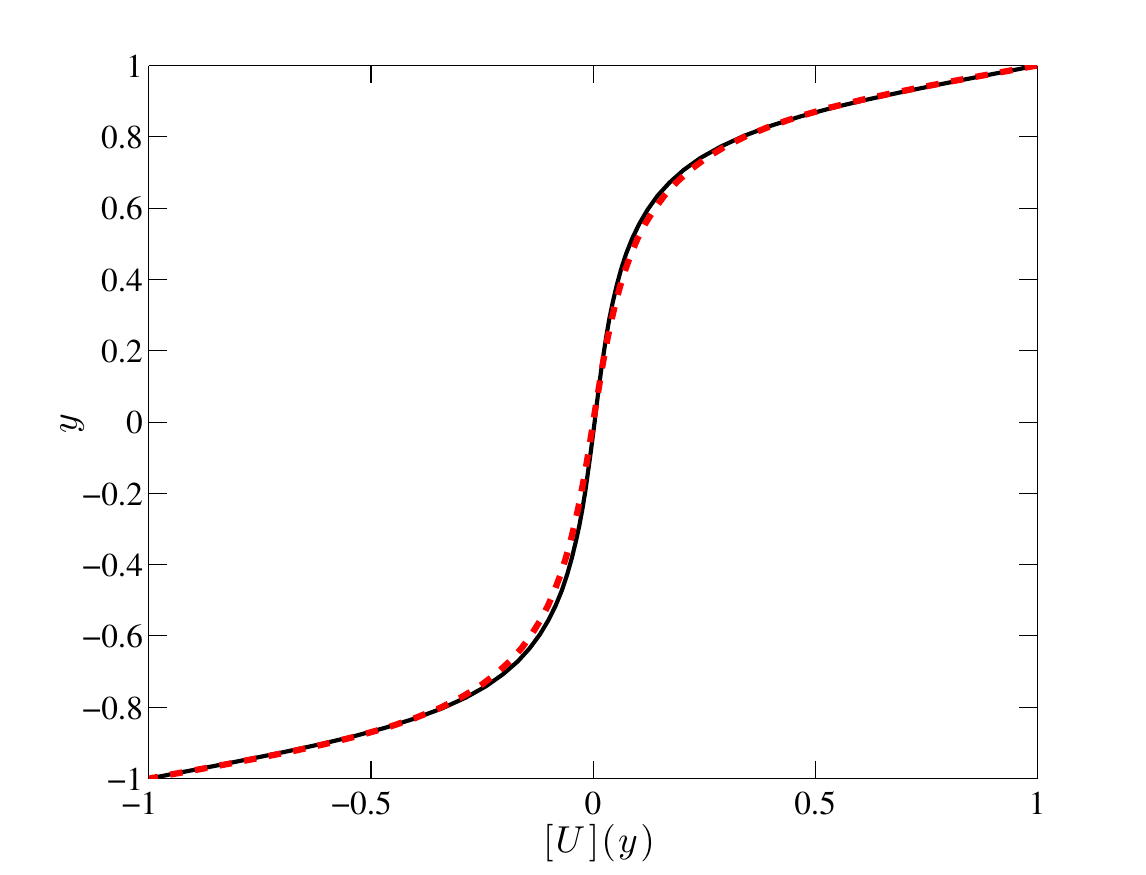}
\caption{
Comparison of the spanwise and time averaged streamwise flow, $[U](y)$, for the self-sustaining state (solid) with the
mean flow obtained from a $128 \times 65 \times 128$ direct numerical simulation (DNS) of Couette turbulence at $R=1000$ in a doubly periodic channel in $x$ and $z$ of length $4 \pi$ in each direction. The S3T self sustaining turbulent state produces on average a
friction velocity based Reynolds number of $R_\tau=64.9$ while the DNS simulation has $R_\tau=66.2$. This Reynolds number indicates the
turbulent production and dissipation and is defined as $R_\tau=u_\tau \delta / \nu$ with $\delta$ the channel half-width and $u_\tau = \sqrt{\nu \left . d [U] /dy \right|_{y = 1}}$
the friction velocity.
This figure demonstrates that the S3T self-sustaining state
produces a streamwise averaged flow profile consistent with simulations of Couette flow turbulence. (Courtesy of V. Thomas)
 }
\label{fig:meanUQ0}
\end{figure}

In analogy with the test function probe used to elucidate the mechanism underlying jet formation in barotropic flow
(cf. section \ref{section:testbar}) we wish
to determine the effect on
the spanwise homogeneous field of turbulence of an
infinitesimal spanwise-dependent mean-flow streak\footnote{The streak component, ${U_s}$, is in general defined as the departure of the streamwise averaged flow ${U}$ from
its spanwise average $[{U}]$, i.e. ${U_s}={U}-[{U}]$.} perturbation $\delta {U_s} (y,z)$ added to the equilibrium flow,
$U^e(y)$. We are particularly interested to determine if distortion
of the turbulence by the perturbation streak results in a positive feedback on the perturbation streak, $\delta {U_s} (y,z)$.
We determine this feedback by
calculating the change $\delta \C_k$ in $\C^e_{k}$ resulting from
the streak perturbation $\delta {U_s} (y,z)$ as in the barotropic example.
The  divergence of the ensemble averaged perturbation Reynolds stresses resulting from this $\delta \C_k$
produce a torque  in the $y-z$ plane inducing a circulation,
according to equation~(\ref{eq:MPSI}), with streamfunction:
\begin{equation}
 \partial_{t} {\delta \Psi} = \Delta_1^{-1} \left[ - (\partial_{yy}-\partial_{zz}) ~\delta\, \overline{{ vw}} - \partial_{yz}~( \delta \,\overline{{w^2}} - \delta \,\overline{{v^2}} \,) \right ]\ .
\label{eq:vddot}
\end{equation}
An example streak perturbations, $\delta { U}_s$, together with vectors of the induced streamwise roll circulation from~(\ref{eq:vddot}),
is shown in Fig.~\ref{fig:benney_random}.
Remarkably, this streak perturbation induces a distortion of the perturbation field resulting in a streamwise roll forcing
configured to amplify the imposed streak perturbation through the lift-up mechanism i.e. positive wall-normal
velocity is collocated with the minimum of the streak and maximum negative wall-normal velocity
is collocated with the maximum of the streak. As a result
this streak perturbation, when imposed on the initially homogeneous field of turbulence,
induces Reynolds stresses driving a roll circulation producing
through lift-up growth of the imposed streak.
This robust Reynolds stress mediated destabilizing feedback process operating on
the streamwise
streak and roll structure
has important implications for both the transition to and the maintenance of turbulence in shear flows.
We will show below that even when the streak structure is highly complex and time-dependent,
as in a turbulent shear flow, the streamwise roll forcing
produced by the perturbation Reynolds stresses remains collocated
so as to amplify  the streak.
Moreover, this property of imposed streaks to  induce, through modification
of the perturbation field, streamwise roll forcing configured to reinforce the imposed
streak provides the mechanism for a streamwise roll and streak plus turbulence cooperative instability in shear flow.
This emergent instability is especially interesting because wall-bounded flows have laminar and turbulent mean velocity profiles
with negative curvature and as a result do not support fast inflectional laminar flow instability
 as a mechanism for robustly transferring energy from the mean flow to the perturbation field as is required in order
 to maintain the turbulent state.
While most streak perturbations organize turbulent Reynolds stresses that do not exactly amplify the streak
that produced them, as is clear in the case of the streak perturbation in Fig.~\ref{fig:benney_random},
if a streak were to organize precisely the perturbation field required for its amplification
then exponential modal growth of this streak and its associated streamwise roll and perturbation fields would result.

\begin{figure}
\centering
\includegraphics[width=19pc,trim = 8mm 0mm 8mm 0mm, clip]{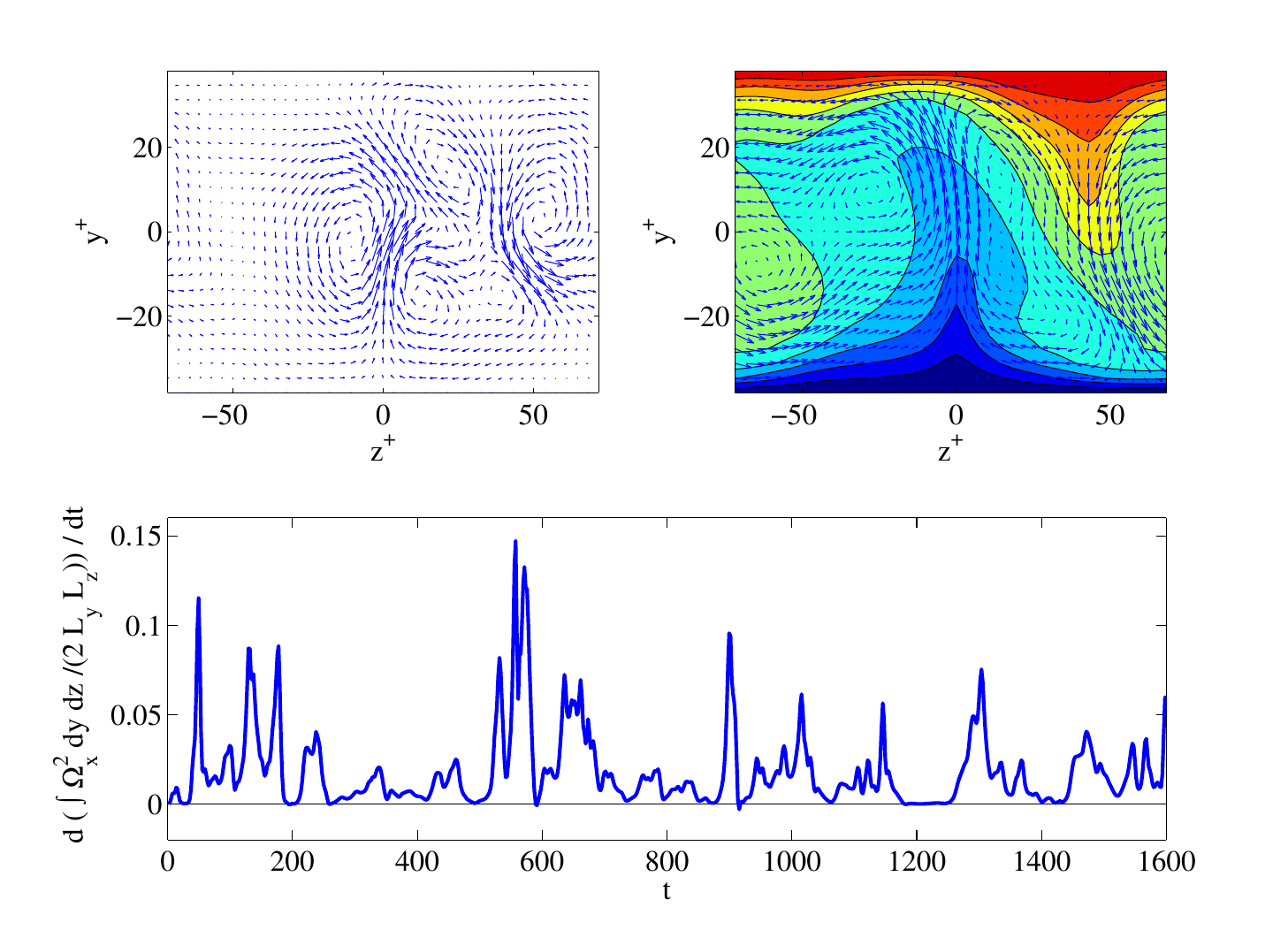}
\caption{
Streamwise roll forcing by perturbation Reynolds stresses in the self-sustaining state with $\epsilon=0$.
Top left: Vectors of instantaneous
cross-stream/spanwise velocity acceleration, $(\dot V, \dot W)$, at time $t=980$.
 Top right: Streamwise roll and streak structure at the same time. Lengths are measured in wall units, $y^+ = R_\tau y $ and $z^+ = R_\tau z $.
 Bottom: Time series of
 streamwise roll forcing as indicated by the rate of change of the average
 square streamwise vorticity.
 It is remarkable that the perturbations, in this highly time-dependent state,
 act to maintain the roll circulation produce,
 not only on average, but also at nearly every instant. Adapted from 
``Dynamics of streamwise rolls and streaks in turbulent wall-bounded shear flow'', by B. F. Farrell and P. J. Ioannou, 2012, 
Journal of Fluid Mechanics, vol. 708, pp. 149-196. \textcopyright\ Cambridge University
Press. Reprinted with permission.}
\label{fig:S-S_torques}
\end{figure}

We determine now the S3T stability of the spanwise homogeneous equilibrium as a function
of excitation amplitude, $\epsilon$, at a fixed Reynolds number, $R$, and show
that exponentially unstable streamwise roll and streak modes
arise in a spanwise independent field of forced turbulence if the perturbation forcing amplitude exceeds a threshold.
The spanwise independent equilibrium is stable for $\epsilon < \epsilon_c $.  At $\epsilon_c$
it becomes structurally unstable, while remaining hydrodynamically
stable. The most unstable eigenfunction, which is shown in Fig.~\ref{fig:8},
consists of a roll circulation with a perfectly collocated streak.
When this eigenfunction is introduced into the S3T system with small amplitude, it grows at first exponentially
at the rate predicted by its eigenvalue and then  asymptotically equilibrates at finite amplitude.
This equilibrium solution, shown in Fig.~\ref{fig:flow8425}, is a steady, finite amplitude streamwise roll and streak.
The bifurcation diagram of the S3T equilibria is shown in Fig.~\ref{fig:HKWbifur} as a function of bifurcation
parameter $\epsilon$.
The finite amplitude streamwise roll and streak equilibria are S3T stable for $\epsilon_c \le \epsilon \le \epsilon_u$.

At  $\epsilon_u$ there is a second bifurcation in which the equilibrium
becomes S3T unstable, while remaining hydrodynamically stable, and
the SSD fails to equilibrate, instead transitioning directly to a time-dependent state.
Remarkably, the time-dependent S3T state that emerges for $\epsilon> \epsilon_u$ self-sustains even when
$\epsilon$ is set to $0$.
This S3T self-sustaining time-dependent state produces realistic turbulence with
mean turbulent profile $[U]$ shown in Fig.~\ref{fig:meanUQ0}.
Moreover, comparison with direct numerical simulations (DNS) verifies that this S3T turbulence is
similar to Navier-Stokes turbulence despite the greatly simplified S3T dynamics underlying it \citep{Farrell-Ioannou-2012-CTR,Constantinou-etal-Madrid-2014,Thomas-etal-2014}.

Remarkably, the S3T self-sustaining state naturally simplifies further by evolving
to a minimal turbulent system
in which the dynamics is supported by the interaction of the roll-streak structures with a perturbation field
comprising a small number of streamwise harmonics (as few as $1$).
This minimal self-sustaining turbulent system, which proceeds naturally from the S3T dynamics,
reveals an underlying self-sustaining process (SSP)  which can be understood with clarity.
The basic ingredient of this SSP is the
robust tendency for streaks to organize the perturbation field so as to produce Reynolds stresses supporting the
streak, via the lift-up mechanism as illustrated in Fig.~\ref{fig:benney_random}.
Although the streak
is strongly fluctuating  in the self-sustaining state,
the tendency of the streak to organize the perturbation field is retained as illustrated in Fig.~\ref{fig:S-S_torques}
in which a snapshot of the streamwise roll and streak is shown together with the associated roll
acceleration, $(\dot V, \dot W)$, 
arising from the perturbation Reynolds stresses (cf. equation~(\ref{eq:vddot})).
The time derivative of the integral square streamwise vorticity, $d/dt ( \int dy\, dz~ \Omega_x^2)$ with
$\Omega_x=W_y-V_z$, provides a measure of the torque produced by the Reynolds stress divergences that
support the roll circulation.
 A times series of this diagnostic is also shown in Fig.~\ref{fig:S-S_torques}.
It is remarkable that the perturbations, in this highly time-dependent state, produce torques that maintain the streamwise roll
not only on average but at nearly every instant.
 As a result, in this self-sustaining state, the streamwise roll is systematically maintained by the
 robust organization of perturbation
Reynolds stress by the time-dependent streak that was identified by SSD analysis using
the S3T system, while the streak is maintained by the streamwise
roll through the lift-up mechanism.  Through the resulting time-dependence of the roll-streak structure
the constraint on instability imposed
by the absence of inflectional instability in the mean flow is
bypassed and the perturbation field is maintained by parametric growth, thus completing the
SSP cycle\footnote{It has been shown that flows that at each time instant satisfy the necessary
for stability Rayleigh condition can still become exponentially unstable if the flow is time
dependent (cf. \cite{Farrell-Ioannou-1999}).}.

We conclude that the dynamics of turbulence in wall-bounded shear flow can be understood at a fundamental level
by using SSD and specifically by exploiting
the direct relation between wall turbulence and
the highly simplified S3T turbulence. Among the results obtained is that the mechanism of turbulence in wall bounded shear flow
is the same roll/streak/perturbation SSP that has been shown to maintain S3T turbulence.

\section{Discussion}

Although turbulence is commonly thought of as being definitional of disorder, the turbulent state in shear flow possesses
an underlying order that is revealed by adopting statistical state variables to characterize the dynamics of the turbulent state.
This fundamental order at the center of
turbulence dynamics has remained incompletely appreciated for lack of a conceptual basis as well as of
methods for analyzing the order underlying statistical states of turbulence.
In this review we have described an approach to understanding emergence of order in turbulence through
adopting the perspective of SSD.
From this perspective order is understood
to arise in turbulence due to
systematic cooperative interaction between large scale structures and the field of small scale turbulence in which these structures are embedded.
Motivating examples of order emergence arise from considering
a field of turbulence of some kind (GFD, MHD, Navier-Stokes) into which a trial perturbation
of small amplitude but large scale is introduced. When introduced, such a structure both alters
the turbulence and is altered by it.  Most such structures are not systematically maintained
by this interaction. But suppose that we were
to continue trying different large scale perturbations until we hit on a perturbation structure that affected the turbulence in just such a manner as to produce Reynolds stresses configured to amplify this structure without changing its form.  Such a structure would naturally grow spontaneously out of the turbulence and
would provide an explanation for the observed emergence of large scale structure in turbulent flow.
This concept of coherent structure emergence through cooperative multi-scale interaction in
turbulence takes analytical form through eigenanalysis of the SSD of
the turbulence linearized about an equilibrium turbulent statistical state. Bifurcations occur in association
with these instabilities as parameters of the problem, such as the amplitude of the turbulence excitation rate,
the damping rate of the flow, or the beta parameter, are varied.
Extensions of these instabilities into the nonlinear regime of the SSD reveal fixed point equilibria
that predict the associated finite amplitude coherent structures e.g. the zonal jets
in the case of barotropic and baroclinic turbulence in planetary atmospheres.
Moreover, these finite amplitude equilibria of the nonlinear SSD have an interpretation that transcends prediction of jet structure.
The statistical state of the turbulence comprises both the mean flow and the perturbations which have mutually adjusted to produce the equilibrium
statistical state
so that these equilibria constitute a closure of the associated turbulence dynamics. The success of SSD in predicting the statistical mean
state of turbulence as S3T equilibria in the systems we have discussed argues constructively that the nonlocal in spectral space interaction between
perturbations and large scale mean flows that is retained in S3T captures
the physical mechanism underlying the maintenance of the turbulent state in these systems.
In addition, this closure is deterministic so
that the physical mechanisms producing the closure are made available for study through analysis of the SSD
underlying them.
A straightforward example of such an insight into closure in turbulence arises in the case of barotropic beta-plane
turbulence in which the observed jet scale in strongly excited turbulence is linked mechanistically through SSD with
Rayleigh's stability criterion; SSD analysis makes the further and associated verifiable prediction of successive
bifurcations to jet structure of smaller wavenumber with increase in turbulence intensity.
This example shows the power of SSD to both predict and provide physical explanation for observed phenomena in
turbulent flows.

Adopting the perspective of SSD also provides new conceptual insights into the dynamics of
turbulence; an example of this is the concept of the dynamical trajectory of a statistical state.
Consider a statistical state equilibrium consisting of a fixed point. An example might be a barotropic
jet together with its supporting turbulence. This fixed point corresponds to a probability density function
that is stationary in state space.
But just as sample state trajectories, which are points in phase space, may converge to a fixed point or
follow a time dependent path through the state space, so also statistical state trajectories may follow
more or less intricate paths in phase space, taking their probability density function along with them. Interpretation of the
dynamics of these statistical state trajectories provides a way to deepen understanding of the dynamic of turbulence.
The successive bifurcations with increase in excitation leading to lower and lower meridional wavenumber for the equilibrium jet
in the example of beta-plane turbulence mentioned above can be viewed as a trajectory of the deterministic SSD with
the bifurcations arising from instability of the evolving statistical state. These trajectories and
instabilities have no analytical counterpart in sample state dynamics.
A familiar example of limit cycle behavior of a statistical state trajectory is provided by the
QBO of the Earth's equatorial stratosphere which exhibits nearly periodic 27
month cycles in phase space. The analytical structure of the associated bifurcation only
exists in the SSD framework \citep{Farrell-Ioannou-2003-structural}. Chaotic statistical state
trajectories are also found in jet dynamics of plasma turbulence \citep{Farrell-Ioannou-2009-plasmas}.  In the case of wall-bounded
shear flow turbulence the statistical state trajectory is also chaotic and the time dependence of this statistical state,
consisting of the streak structure and the associated perturbations it supports, which results from their cooperative
interaction, is a fundamental component of the dynamical mechanism underlying the
SSP maintaining the turbulence. This is because the
turbulence is maintained by perturbations that result from parametric interaction with the streak, the
time dependence of which is in turn maintained by interaction with the perturbations.  This
parametric process underlies energy transfer from the inflectionally stable forced mean flow that is required to maintain
the turbulent perturbation variance in wall-bounded shear flows.
This parametric growth process is intrinsically a property of the SSD and this
cooperative parametric process can only be understood through analysis of the dynamics of the statistical state of the turbulence.

Viewing turbulence from the perspective of SSD has proven to be remarkably tractable and to provide a richness of analysis and concept that has already allowed progress in a number of areas and holds promise of continuing insight into the nature of turbulence.


\ifthenelse{\boolean{dc}}
{}
{\clearpage}

\begin{appendix}[A]
\addcontentsline{toc}{section}{\protect\numberline{}Appendices}
\section*{\begin{center}The homogeneous equilibrium covariance\end{center}\label{appA}}

We prove that the equilibrium covariance (\ref{eq:Ceq}) under homogeneous stochastic excitation of a channel with periodic boundary conditions
produces zero mean vorticity fluxes, $\overline{v' q'} =0$. This is required in order for $U^e=0$ to be an
S3T equilibrium with perturbations with covariance (\ref{eq:Ceq}). We provide a proof that
uses technical arguments that are useful for exploring the properties
of discrete S3T dynamics and stability in periodic channels  (cf. \cite{Bakas-Ioannou-2011}).
 We use the property that in the matrix formulation
 of S3T in a periodic channel, in which the fields have been discretized on a grid,
 all matrices that correspond to homogeneous continuous operators, as well as all covariances
 of homogeneous fields, are circulant, i.e. each row is a cyclic shift of the row above it. They are circulant
because they commute with the matrix of spatial shifts on the grid lattice. Circulant matrices commute with each other and their common eigenbasis
is a unitary matrix consisting of harmonics.
 Consequently the forcing
 covariance at wavenumber $k$  is analyzed in Fourier components in $y$ as
 $Q_{ k \alpha \beta}= \sum_{l} {\hat Q}_{k l} e^{\i l (y_\alpha - y_\beta)}$, with $l$ the $y$ wavenumber.
 Moreover, the Fourier coefficients $ \hat{Q}_{k l} $ must be real and non-negative coefficients in order to assure that  $Q_{k \alpha \beta}$, for each $k$, is positive definite,
 Hermitian and a covariance of a homogeneous field
 (this statement is the content of Bochner's theorem).
 Because for each $k$ we have
 \begin{align*}
 \sum_{m=1}^n \Delta_{k \alpha m }^{-1} Q_{k m \beta }
 &=  \sum_{l} \hat{Q}_{k l} \sum_{m=1}^n \Delta_{k \alpha m }^{-1} e^{\i l (y_m - y_\beta)} \nonumber \\
 &=  -\sum_{l} \hat{Q}_{k l}  \frac{ e^{\i l (y_\alpha - y_\beta)}}{k^2 +l^2}
 \end{align*}
we obtain that the diagonal elements of  $\sum_{m=1}^n \Delta_{k \alpha m }^{-1} C_{k m \beta }^e$
are equal and real and therefore the vorticity flux associated with the equilibrium covariance, which is proportional to the imaginary part of the diagonal elements of this matrix,
vanishes.

\end{appendix}




\end{document}